\documentclass[aps,prl,reprint,superscriptaddress,longbibliography,nofootinbib,floatfix]{revtex4-2}

\usepackage[T1]{fontenc}
\usepackage[utf8]{inputenc}
\usepackage{amsmath,amssymb,amsthm,mathtools}
\usepackage{bm}
\usepackage{graphicx}
\usepackage{booktabs}
\usepackage{multirow}
\usepackage{array}
\usepackage{braket}
\usepackage{xcolor}
\usepackage{hyperref}
\usepackage{physics}

\hypersetup{
  colorlinks=true,
  linkcolor=blue,
  citecolor=blue,
  urlcolor=blue
}

\newcommand{\psucc}{p_{\mathrm{succ}}}

\newcommand{\NOON}{\mathrm{NOON}}

\newcommand{\prlheading}[2][]{%
  \par\addvspace{0.65\baselineskip}%
  \noindent\emph{#2---}%
  \if\relax\detokenize{#1}\relax\else\phantomsection\label{#1}\fi
  \ %
  \ignorespaces
}

\newcounter{supsec}
\newcounter{supsubsec}[supsec]

\renewcommand{\thesupsec}{S.\arabic{supsec}}
\renewcommand{\thesupsubsec}{\thesupsec.\arabic{supsubsec}}

\newcommand{\supsection}[1]{%
    \refstepcounter{supsec}%
    \setcounter{supsubsec}{0}%
    \setcounter{equation}{0}%
    \renewcommand{\theequation}{\thesupsec.\arabic{equation}}%
    \renewcommand{\theHequation}{%
        supplement.\arabic{supsec}.0.\arabic{equation}%
    }%
    \section*{\thesupsec\quad #1}%
}

\newcommand{\supsubsection}[1]{%
    \refstepcounter{supsubsec}%
    \setcounter{equation}{0}%
    \renewcommand{\theequation}{\thesupsubsec.\arabic{equation}}%
    \renewcommand{\theHequation}{%
        supplement.\arabic{supsec}.\arabic{supsubsec}.\arabic{equation}%
    }%
    \subsection*{\thesupsubsec\quad #1}%
}

\begin{document}

\title{Automated discovery of high-probability heralded schemes for path-entangled states}

\author{Marcello Armezzani}
\email{marcello.armezzani@uni-tuebingen.de}
\affiliation{Machine Learning in Science Cluster, Department of Computer Science, Faculty of Science, University of T\"ubingen, Germany}

\author{Colin P. Lualdi}
\email{clualdi2@illinois.edu}
\affiliation{Materials Research Laboratory, The Grainger College of Engineering, University of Illinois Urbana-Champaign, Urbana, IL, USA}
\affiliation{Illinois Quantum Information Science and Technology Center, The Grainger College of Engineering, University of Illinois Urbana-Champaign, Urbana, IL, USA}
\affiliation{Department of Physics, The Grainger College of Engineering, University of Illinois Urbana-Champaign, Urbana, IL, USA}

\author{Xuemei Gu}
\email{xuemei.gu@uni-jena.de}
\affiliation{Institut für Festkörpertheorie und Optik, Friedrich-Schiller-Universität Jena, Jena, Germany}

\author{Paul G. Kwiat}
\email{kwiat@illinois.edu}
\affiliation{Materials Research Laboratory, The Grainger College of Engineering, University of Illinois Urbana-Champaign, Urbana, IL, USA}
\affiliation{Illinois Quantum Information Science and Technology Center, The Grainger College of Engineering, University of Illinois Urbana-Champaign, Urbana, IL, USA}
\affiliation{Department of Physics, The Grainger College of Engineering, University of Illinois Urbana-Champaign, Urbana, IL, USA}

\author{Mario Krenn}
\email{mario.krenn@uni-tuebingen.de}
\affiliation{Machine Learning in Science Cluster, Department of Computer Science, Faculty of Science, University of T\"ubingen, Germany}

\date{July 28, 2026}

\begin{abstract}
Entangled states of light lie at the heart of photonic quantum technologies, from distributed quantum communication to quantum-enhanced measurement and information processing. Their practical generation, however, remains constrained by the weak interactions between photons, which make the deterministic assembly of large multiphoton entangled states a central challenge in quantum optics. In this work, we use AI techniques to discover heralded linear-optical schemes for path-entangled states and show that the resulting solutions can be elevated from individual circuits to a new scalable family. This family contains previously known constructions as special cases while generally providing exponential and super-exponential improvements over those, and its extension to broader classes of target states shows how automated discovery can reveal transferable physical understanding. By presenting compact experimental proposals for large path-entangled states, our results provide both a theoretical advance in photonic heralding and a route towards a substantial leap in experimentally accessible multiphoton entanglement.
\end{abstract}

\maketitle

\prlheading[sec:introduction]{Introduction}
Entangled states of light are central resources for quantum technologies, with applications ranging from quantum communication and computation to sensing and tests of fundamental physics \cite{Prevedel:07, Pan2012MultiphotonEntanglement, Flamini2019PhotonicQIPReview, Couteau2023SinglePhotonsCommunicationComputing, Couteau2023SinglePhotonsMetrologyFoundations}. However, the generation of such states remains fundamentally constrained: in the absence of strong deterministic photon-photon interactions, linear optics alone generally cannot produce arbitrary multiphoton entanglement on demand \cite{AspuruGuzik2012PhotonicQuantumSimulators, OBrien2009PhotonicQuantumTechnologies}. As a result, many critical photonic resources are generated probabilistically, either by post-selection or by heralding. The latter is especially important because it certifies the successful preparation of a desired state without consuming the state itself. Instead, auxiliary photons are measured in ancillary modes, and specific detection patterns announce the presence of the target state in the remaining modes. Combined with switchable quantum memories, this would enable periodic or on-demand preparation of the desired states, making heralded generation a natural route toward scalable photonic quantum technologies \cite{forbes2025heralded}.

Among the paradigmatic examples of multiphoton entanglement are path-entangled states. In their simplest and most widely studied form, two-mode NOON states are defined as
\begin{equation}
\ket{\NOON_N}=\frac{\ket{N,0}+e^{i\phi}\ket{0,N}}{\sqrt{2}},
\label{eq:noon}
\end{equation}
where $N$ is the total photon number and $\phi$ is an arbitrary relative phase between the two terms. These states have played a prominent role in quantum metrology, imaging, and lithography because of their enhanced phase sensitivity in the absence of loss~\cite{Dowling2008,giovannetti2011advances,Pezze2018}. They also represent a fundamental and peculiar form of entanglement alongside Bell and GHZ states \cite{forbes2025heralded}.

A substantial body of work has established foundational heralding strategies for NOON states and related resources, including symmetric multiport constructions, vacuum- or photodetection-assisted schemes, and feed-forward-based approaches~\cite{Zou2002,Kok2002,Pryde2003,Cable2007,mccusker2009efficient}. However, the landscape rapidly becomes more intricate as the target photon number increases, and the challenge is no longer just to parameterize a universal interferometer, but to identify input structures and heralding mechanisms that support exact cancellation of unwanted amplitudes with high success probability.

This challenge naturally suggests automated discovery. There is growing evidence that algorithmic design can uncover genuinely new quantum experiments rather than merely reproducing known intuition~\cite{Krenn2016,Krenn2020ComputerInspired, arlt2026automated, gubarev2020improved, fldzhyan2021compact, krenn2023artificial}. Beside providing clear improvements in the efficiency and fidelity of state generation, the main conceptual question is whether such automated searches can be turned into \emph{physical understanding}~\cite{krenn2022scientific}: do they reveal new mechanisms, new general families, and new scaling laws?

In this work we answer this question affirmatively. Using a large-scale, fast automated search framework we discover a new, exact, and scalable family of circuits for generating states of the form shown in Eq.~\eqref{eq:noon}. This family reduces to known schemes for specific choices of parameters, but generally offers exponential and super-exponential enhancements in the success probability. This family also proves effective for the generation of \emph{multi-mode} $\NOON$ states, improving over the best linear optical families~\cite{Zhang2018MultimodeNOON}. We also report the discovery of improved solutions that are not part of the new family, and extend the same automated strategy to the so-called $m,m'$ path-entangled states~\cite{Huver2008}, obtaining passive linear-optical schemes that outperform the previously known nonlinear family for the same targets~\cite{Glasser2008}. Finally, we address the experimental feasibility of our new solutions and propose future developments.

\prlheading[sec:family]{Automated discovery of a modular comb family}
We build a universal multiport interferometer \cite{Reck1994, Clements2016} in a fast JAX \cite{Bradbury2018JAX} simulator, which we then optimize via stochastic gradient descent methods. We choose an input product state and an output target state; the latter exits in undetected modes while photon-number-resolving (PNR) detectors monitor the remaining (heralding) modes. From each simulated experiment we extract a heralding probability $\psucc$ and a fidelity $F$ with respect to the target state, and we update the circuit to maximize both. Details are provided in Supplemental Material \ref{app:optimization}. 

\begin{figure}[tbp]
\centering
\includegraphics[width=0.96\columnwidth]{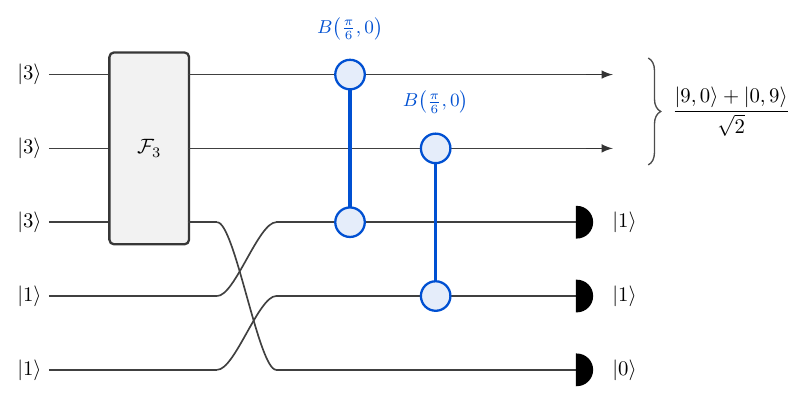}
\caption{\label{fig:noon9_family}
Compact $\NOON_9$ heralded circuit obtained from the input $\ket{3,3,3,1,1}$, which is the optimal choice within the new modular comb family. $\mathcal{F}_3$ is the Fourier multiport transformation acting on the bunched input photons and $B(\theta, \phi)$ is the beam splitter. See Supplemental Material \ref{app:noon9} for details.}
\end{figure}

Searching over many input and target states, and carefully analyzing the multitude of results produced, a particularly interesting general family emerges. Its first newly discovered member is the heralded circuit for \(\NOON_9\), shown in Fig.~\ref{fig:noon9_family}. 
\noindent For this specific case, the success probability increases by almost \(1440\%\) with respect to the best existing scheme \cite{Pryde2003}. As additional benefits, the circuit requires fewer detectors and no longer relies solely on vacuum heralding---heralding on a detector \emph{not} firing---which is known to be an unreliable experimental condition \cite{PhysRevA.104.033717}.

The general mechanism of the family can be described starting from 
\begin{equation}
    N=\sum_{\ell=1}^L r_\ell m_\ell,
    \label{eq:N}
\end{equation}
where $r_\ell \geq 2$ and $m_\ell \geq 1$. The first part of the scheme takes an input state composed of $L$ packets $\bigotimes_{\ell=1}^L
    \ket{m_\ell}^{\otimes r_\ell}$ and propagates it through a multiport that leaves the two target modes populated by the modular comb 
\begin{equation*}
    \ket{N-q,q},
    \quad
      q \in
   Q,
\end{equation*}
where $Q$ denotes the actual non-zero photon-number support of the comb. All the unwanted middle terms between $\ket{N,0}$ and $ \ket{0,N}$ can be discarded through single-photon Fock filters, using a state $\ket{1,1}^{\otimes|B|}$, where $B=
\left\{
\min(q,N-q):
q\in Q,\ q\neq 0,N
\right\}$ is the set of distinct symmetric filters. One mixes the target mode with an ancillary single photon on a beam splitter, and then keeps only the events in which one photon is detected again in the ancillary output. The heralded output is exactly the desired $\NOON$ state in Eq. \eqref{eq:noon} with a success probability of
\begin{equation}
    \psucc
    =
    2
    \frac{N!}{N^N}
    \prod_{\ell=1}^{L}
    \left(
        \frac{m_\ell^{m_\ell}}{m_\ell!}
    \right)^{r_\ell}
    \prod_{b\in B}
    \left(
        \frac{b}{b+1}
    \right)^N
    \left(
        \frac{N-b}{b+1}
    \right)^2.
    \label{eq:fourier_comb_family_probability_text}
\end{equation}

\noindent Details on the derivation can be found in \ref{app:explicit derivation}. A remarkable aspect of Eq.~\eqref{eq:fourier_comb_family_probability_text} is that it contains, as limiting cases, the two main passive linear-optical families known before this work. Setting $L=1$ gives $N=rm$. Choosing $r=N$, $m=1$ produces no interior sectors to filter, and reduces the construction to the Pryde and White (PW) multiport family~\cite{Pryde2003}. For even $N$, choosing $r=2$, $m=N/2$ gives the even-$N$ family of Zou, Pahlke, and Mathis (ZPM)~\cite{Zou2002}. These two limiting cases represent the previous state of the art for passive linear-optical $\NOON$-state generation~\cite{forbes2025heralded}.

\begin{figure}[tbp]
\centering
\includegraphics[width=1\linewidth]{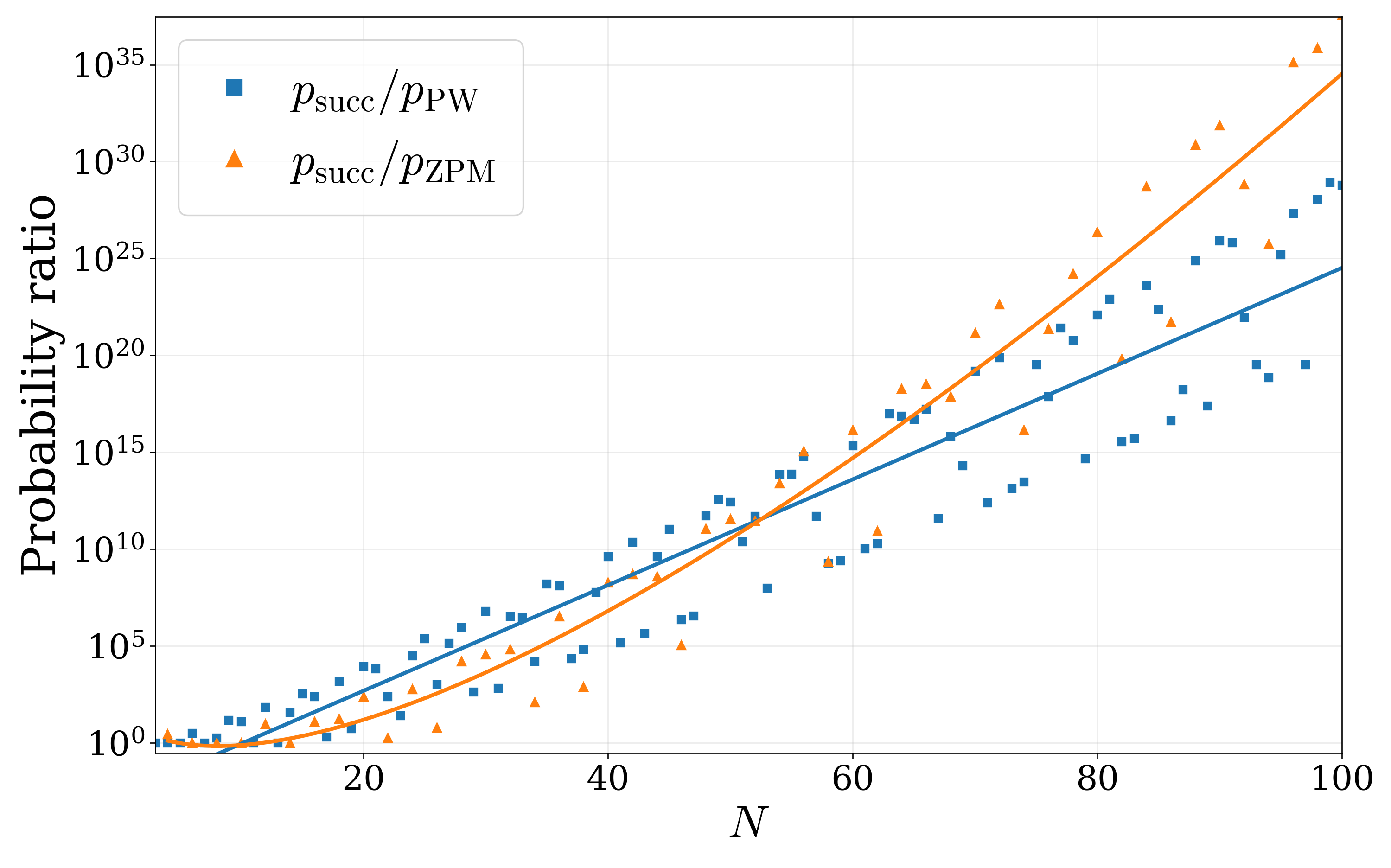}
\caption{Ratios between the success probability of the modular-comb family and the probabilities of the PW and ZPM schemes as $N$ grows. The points are exact optimized values obtained by choosing the best decomposition and filter set for each $N$. The solid curves are finite-$N$ fits with the asymptotic forms derived in the Supplemental Material \ref{app:asymptotic_two_modes}.}
\label{fig:improvement}
\end{figure}

Furthermore, the nontrivial branches of the modular-comb family improve asymptotically over both of these known schemes. In Supplemental Material \ref{app:asymptotic_two_modes}, we show that explicit fixed-shape branches already satisfy
\begin{equation}
\frac{\psucc}{p_{\rm PW}}
\geq
10^{c_{\rm PW}N+O(1)},
\frac{\psucc}{p_{\rm ZPM}}
\geq
10^{\frac{N}{2}\log_{10}N+O(N)} ,
\label{eq:scaling}
\end{equation}
where $c_{\rm PW} > 0$. Thus, the improvement is at least exponential with respect to PW and super-exponential with respect to ZPM. The exact optimized envelope is not monotonic because the best integer decomposition of $N$ and the optimal filter set depends on the value of $N$. In Fig.~\ref{fig:improvement}, we show the exact optimized ratios together with finite-$N$ fits of the forms implied by Eq.~\eqref{eq:scaling}.

\begin{figure}[tbp]
    \centering
    \includegraphics[width=1\linewidth]{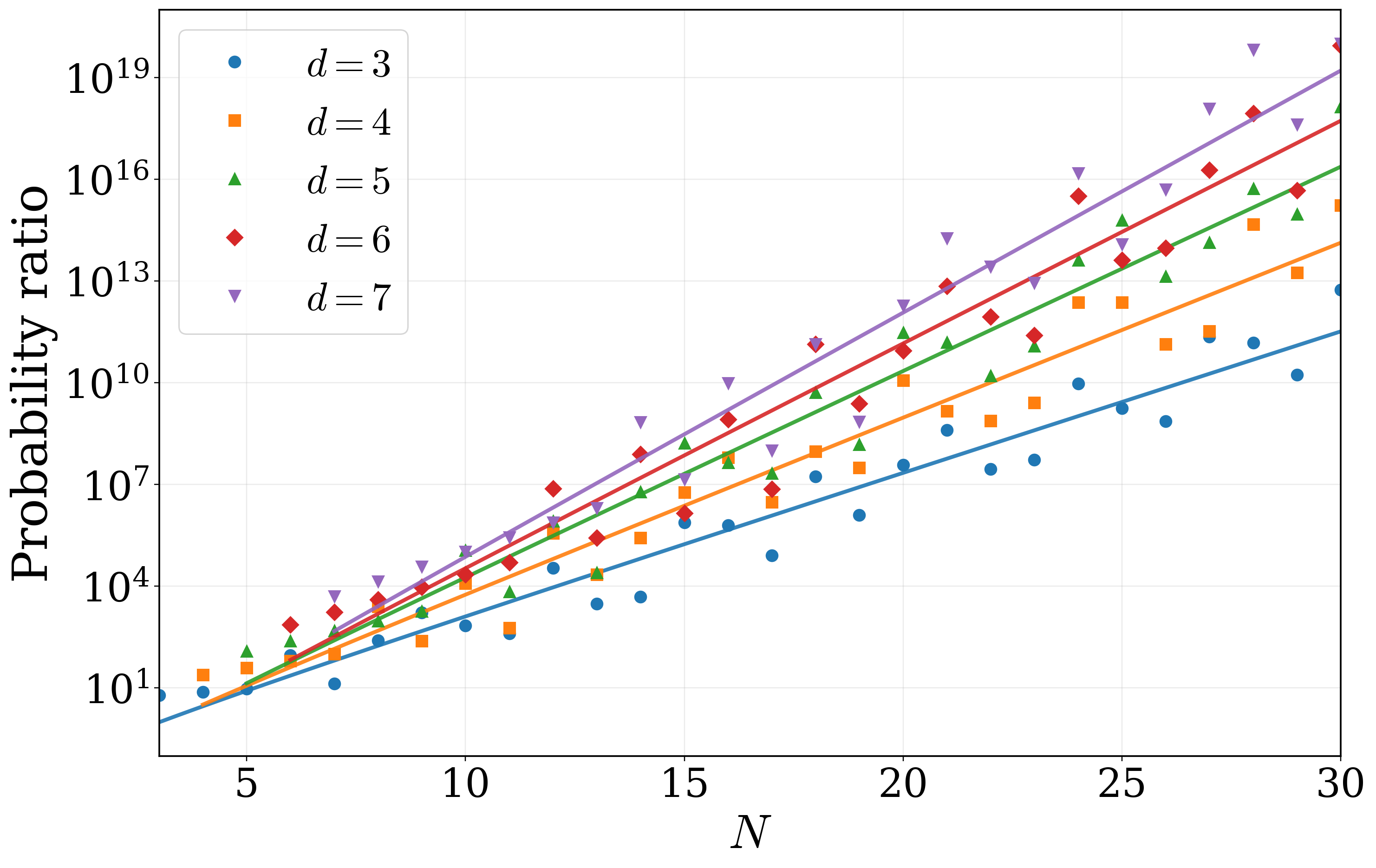}
    \caption{Ratio between the success probability of the modular-comb extension to multimode $\NOON$ states and the Zhang--Chan scheme as $N$ grows, for $d=3,\ldots,7$. The comparison is limited to $N\geq d$, since the modular comb family requires $r_\ell \geq d$. The points are exact optimized values under the common universal
filtering prescription, obtained by choosing the best decomposition for each pair $(N,d)$. The solid curves are finite-$N$ fits of the asymptotic form derived in Supplemental Material \ref{app:asymptotic_multi_modes}.}
    \label{fig:multi_mode_improvement}
\end{figure}

\prlheading[sec:multi-mode]{Understanding by generalization to the multi-mode $\NOON$ case}
Scientific understanding is often revealed by the ability to generalize a mechanism beyond the particular case in which it was first identified~\cite{de2017understanding}. We therefore test whether the modular-comb idea that emerged from the automated discovery for two-mode $\NOON$ states can be extended to multi-mode $\NOON$ states, which are particularly interesting for the simultaneous estimation of multiple parameters~\cite{humphreys2013quantum,yue2014quantum}. We consider $N$ written as in Eq. \eqref{eq:N}, with the requirement $r_\ell \geq d$, where $d$ is the number of modes of the target multi-mode $\NOON$ state. As before, we send $L$ packets through a multiport, which this time populates the modular comb (details in Supplemental Material \ref{app:derivation_multi_mode})
\begin{equation*}
\ket{N-|\mathbf q|,q_1,\ldots,q_{d-1}},
\qquad
|\mathbf q|=\sum_{\mu=1}^{d-1}q_\mu .
\end{equation*}
Once again the single-photon Fock filters cancel out all the unwanted terms, leaving us with $\ket{N,0,\ldots,0},\ket{0,N,\ldots,0},\ldots,\ket{0,\ldots,N,0},\ket{0,\ldots,0,N}.$
This family improves over the best-performing linear-optical scheme presented by Zhang and Chan in~\cite{Zhang2018MultimodeNOON}. In Supplemental Material~\ref{app:asymptotic_multi_modes}, we show that, for fixed $d$, every fixed-shape branch of the modular-comb
family satisfies
\begin{equation}
    \frac{\psucc^{(d)}}{p_{\rm ZC}^{(d)}}
    \geq
    N^{1/2}10^{c_{\rm ZC}^{(d)}N+O(1)},
    \qquad
    c_{\rm ZC}^{(d)}>0.
    \label{eq:multimode_asymptotic_ratio}
\end{equation}
Here, $c_{\rm ZC}^{(d)}$ depends on the chosen fixed-shape branch. In Fig.~\ref{fig:multi_mode_improvement}, we plot the exact optimized ratios under the common universal filtering prescription. together with finite-$N$ fits to the regular-branch form $N^{1/2}10^{a_dN+b_d}$.

\prlheading[sec:additionalschemes]{Additional schemes}
While the modular-comb family significantly improves on the previous state of the art, the new scheme is not necessarily optimal for all values of $N$. Indeed, the automated discovery process reveals other superior solutions for specific target states that cannot be forced inside the same picture. In addition, to illustrate the flexibility of our approach, we also automate the discovery of the so-called $m,m'$ states \cite{Huver2008}
\begin{equation*}
    \frac{\ket{m,m'}+\ket{m',m}}{\sqrt{2}},
\end{equation*}
originally introduced to make $\NOON$ states more robust to loss. It is known in fact that advantages offered by $\NOON$ states rapidly decrease when loss is not negligible \cite{PhysRevA.75.053805}. We summarize few examples of our findings in Table \ref{tab:additional_schemes}. These examples, while not exhaustive, serve as evidence that it is certainly possible to discover improved solutions that do not belong to general families; this was also recently shown in \cite{singh2025heralded} for the three-mode, two-photon $\NOON$ state: $\left(\ket{2,0,0} + e^{i\alpha_1}\ket{0,2,0} + e^{i\alpha_2}\ket{0,0,2}\right)/\sqrt{3}$.

\begin{table*}[!t]
    \centering
    \scriptsize
    \setlength{\tabcolsep}{3.2pt}
    \renewcommand{\arraystretch}{1.22}
    \caption{Examples of other discovered schemes that are not part of the newly discovered modular comb family. Automated discovery unlocks solutions that, beside achieving better success probabilities, reduce the required number of heralding modes (i.e., detectors) and eliminate complete vacuum conditions at the detectors. The explicit circuit of the new $\NOON_8$ solution is shown in Fig. \ref{fig:noon-8}. The other schemes are explicitly shown in Supplemental Material \ref{app:additional schemes}.}
    \label{tab:additional_schemes}
    \begin{tabular*}{\textwidth}{@{\extracolsep{\fill}}|c|c|c|c|c|c|}
        \hline
        Target state
        & Best known probability
        & New probability
        & Gain
        & Detectors condition
        & Source requirement \\
        \hline

        $\displaystyle \frac{|5,0\rangle+|0,5\rangle}{\sqrt{2}}$
        &
        $\sim 7.68\%$~\cite{Pryde2003}
        &
        $\sim 9.17\%$ 
        &
        $\sim 19\%$
        &
        \begin{tabular}[c]{@{}c@{}}
            old: $\ket{000}$ \\
            new: $\ket{10}$
        \end{tabular}
        &
        \begin{tabular}[c]{@{}c@{}}
            old: $|1\rangle$ \\
            new: $|2\rangle$
        \end{tabular}
        \\
        \hline

        $\displaystyle \frac{|7,0\rangle+|0,7\rangle}{\sqrt{2}}$
        &
        $\sim 1.22\%$~\cite{Pryde2003}
        &
        $\sim 3.13\%$ 
        &
        $\sim 155\%$
        &
        \begin{tabular}[c]{@{}c@{}}
            old: $\ket{00000}$ \\
            new: $\ket{100}$
        \end{tabular}
        &
        \begin{tabular}[c]{@{}c@{}}
            old: $|1\rangle$ \\
            new: $|2\rangle$
        \end{tabular}
        \\
        \hline

        $\displaystyle \frac{|8,0\rangle-|0,8\rangle}{\sqrt{2}}$
        &
        $\sim 0.92\%$~\cite{Zou2002}
        &
        $\sim 5.44\%$ 
        &
        $\sim 494\%$
        &
        \begin{tabular}[c]{@{}c@{}}
            old: $\ket{1111}$ \\
            new: $\ket{11}$
        \end{tabular}
        &
        \begin{tabular}[c]{@{}c@{}}
            old: $|4\rangle$ \\
            new: $|3\rangle$
        \end{tabular}
        \\
        \hline

        $\displaystyle \frac{|300\rangle+|030\rangle+|003\rangle}{\sqrt{3}}$
        &
        \begin{tabular}[c]{@{}c@{}}
            $\sim 0.69\%$~\cite{Zhang2018MultimodeNOON} \\
            $\sim 4.17\%$ comb family
        \end{tabular}
        &
        $\sim 14.71\%$ 
        &
        \begin{tabular}[c]{@{}c@{}}
            $\sim2018\%$ over old \\
            $\sim 253\%$ over comb
        \end{tabular}
        &
        \begin{tabular}[c]{@{}c@{}}
            old/comb: $\ket{111}$ \\
            new: $\ket{10}$
        \end{tabular}
        &
        \begin{tabular}[c]{@{}c@{}}
            old: $|3\rangle$ \\
            new/comb: $|1\rangle$
        \end{tabular}
        \\

        \hline

        $\displaystyle \frac{|L,L-2\rangle+|L-2,L\rangle}{\sqrt{2}}$
        &
        $\scriptstyle
        \frac{8(L-1)}{L^3}
        \left(\frac{L-2}{L}\right)^{2L-4}$~\cite{Glasser2008}
        &
        $\scriptstyle
        \frac{1}{2}
        \left(1-\frac{1}{L}\right)^{2L-3}$
        &
        \begin{tabular}[c]{@{}c@{}}
            quadratic \\
            asymptotic gain
        \end{tabular}
        &
        \begin{tabular}[c]{@{}c@{}}
            old: $\ket{L-2,L-2}$ \\
            new: $\ket{11}$
        \end{tabular}
        &
        \begin{tabular}[c]{@{}c@{}}
            old: increasing \\ nonlinearities \\
            new: $|L-1\rangle$
        \end{tabular}
        \\
        \hline
    \end{tabular*}
\end{table*}

\begin{figure}[tbp]
    \centering
    \includegraphics[width=1\linewidth]{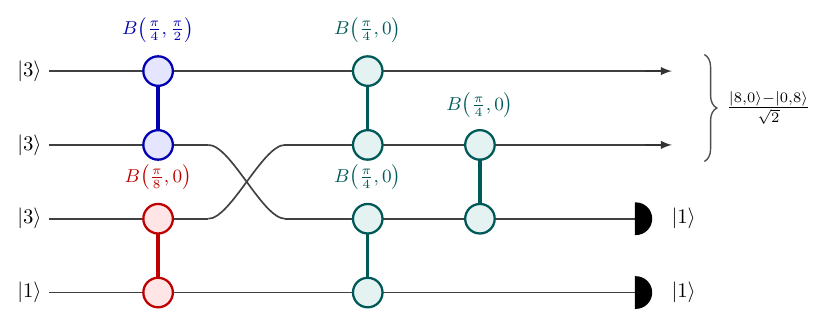}
    \caption{Compact optimized $\NOON_8$ circuit obtained from the input $\ket{3,3,3,1}$ with heralding pattern $\ket{1,1}$. The circuit produces $(\ket{8,0}-\ket{0,8})/\sqrt{2}$. Compared with the ZPM construction, it uses maximum input occupation 3 instead of 4, requires two heralding detectors instead of four, and increases the success probability by approximately 500$\%$.}
    \label{fig:noon-8}
\end{figure}

\begin{figure*}[!t]
    \centering
    \includegraphics[width=0.8\linewidth]{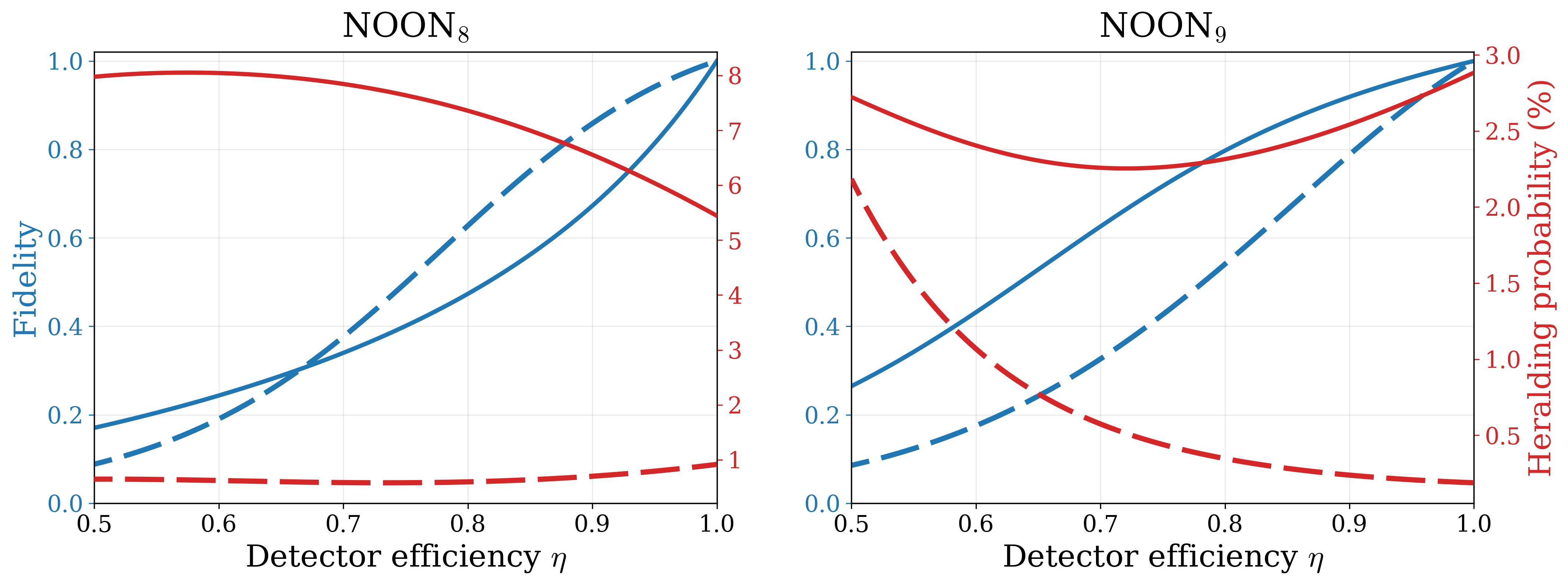}
    \caption{We compare fidelity and heralding probability for the schemes in Figs. \ref{fig:noon9_family} and \ref{fig:noon-8} as well as the original PW and ZPM proposals as the detectors' efficiency $\eta$ vary. The heralding probability is the probability of producing the heralding click pattern at the detectors. The solid curves correspond to the new discovered circuits, while the dashed ones indicate the performance of the previous schemes.}
    \label{fig:experimental_feasibility}
\end{figure*}

\prlheading[sec:feasibility]{Experimental feasibility}
We now address the main experimental challenges that our proposed schemes would face. To do so, we analyze as examples the $\NOON_9$ and $\NOON_8$ circuits (Figs. \ref{fig:noon9_family} and \ref{fig:noon-8}, respectively). We first consider the consequences of inefficient heralding detectors. To model a detector with efficiency $\eta$, we incorporate a beam splitter with transmissivity $\eta$ and reflectivity $1-\eta$. The results are shown in Fig. \ref{fig:experimental_feasibility}, where we compare the new $\NOON_9$ and $\NOON_8$ circuits with the original PW and ZPM proposals, respectively. In the case of the new $\NOON_9$ scheme, both the fidelity and the heralding probability consistently exceed those of PW as the detector efficiency decreases. The fidelity also decreases more slowly with the new scheme. These results are due to the replacement of the full vacuum heralding condition $\ket{0000000}$ with the more reliable $\ket{110}$. For $\NOON_8$, we observe that the stricter ZPM $\ket{1111}$ heralding condition---compared to $\ket{11}$ for our new scheme (see Table \ref{tab:additional_schemes})---better preserves the fidelity under decreasing efficiency, albeit at the cost of a greatly reduced heralding probability.

Experimental implementation of reliable heralding benefits from many options for efficient detectors, ranging from transition-edge sensors (e.g., \cite{morais2024precisely, eaton2024resolution}) to superconducting nanowire single-photon detectors  (e.g., \cite{davis2022improved, stasi2025enhanced}), with various experimental trade-offs involving efficiency, photon-number-resolving capabilities, detection rate, and implementation complexity. Given the availability of detectors with up to at least 98\% efficiency \cite{reddy2020quantum,alexander2025manufacturable}, some with excellent photon-number resolution, we conclude that achieving high fidelity and heralding probability is possible. 

Concerning the input states, both of the new schemes require  $\ket{1}$ and $\ket{3}$. Since the complexity of producing a Fock state generally increases with its size, our $\NOON_8$ result is advantageous compared to the original ZPM proposal, which requires $\ket{4}$ instead of $\ket{3}$. Interestingly, the same does not apply to our $\NOON_9$ result: the original PW scheme requires only $\ket{1}$ states.

There has been significant progress in recent years with regard to generating these input states. For example, state-of-the-art single-photon sources can currently achieve success probabilities around 70\% \cite{kaneda2019high, ding2025high}. While the efficient generation of larger Fock states such as $\ket{3}$ has yet to be demonstrated, multiple theoretical studies show that a ``repeated addition'' scheme can realize success probabilities in the tens of percent range with optimistic but realistic experimental parameters \cite{mccusker2009efficient, glebov2014photon}. Although realizing such devices remains a work in progress \cite{engelkemeier2021climb}, such a protocol is far more promising than established methods based on single-pass spontaneous parametric down-conversion sources (e.g.,  \cite{waks2006Gen, cooper2013exp, engelkemeier2021climb}), which offer success probabilities that are lower by orders of magnitude. We also note that generating multiphoton states with a quantum-dot source is an emerging option \cite{wu2026purcell, karli2025passive}.  

Finally, the small number of modes (and beamsplitters -- five for each scheme) makes circuit construction straightforward. Loss can be minimized by leveraging free-space bulk optics: commercially available beamsplitters can offer $<$1\% loss \cite{thorlabs_uvfs_bs} and 97\% coupling efficiency into single-mode fiber has been reported in the literature \cite{zhong2018twelve}. Even on-chip implementation may be possible given recent advances in low-loss SiN waveguide components featuring $\sim$0.01\% splitter loss and up to $\sim$99\% chip-to-fiber coupling \cite{alexander2025manufacturable}. Altogether, the experimental feasibility of efficient detectors, sources, and circuits lead us to believe that high-rate generation of $\NOON_8$ and $\NOON_9$ states based on our newly discovered schemes is attainable in the near future.

\prlheading[sec:Discussion]{Outlook}
In this work we presented a new exact family for the heralding of $\NOON$ states that improves exponentially and super-exponentially over existing schemes as $N$ grows. Additionally, we proved that the same family can be extended to multi-mode $\NOON$ states, and presented examples of superior schemes that do not belong to the new family. 

\textbf{Experimental implementations} -- We discussed experimental advantages offered by the newly discovered $\NOON_8$ and $\NOON_9$ circuits compared to existing schemes: the former requires smaller Fock states as inputs and the latter offers more favorable fidelity scaling under decreasing detector efficiency by virtue of minimizing vacuum heralding. These findings make these circuits strong candidates for near-future implementation. 

\textbf{Towards higher-photon-number heralding} -- Notable heralded-entanglement experiments \cite{wagenknecht2010experimental, barz2010heralded, matthews2011heralding, hamel2014direct, chen2024heralded, cao2024photonic} have reported a maximum target-state size of four photons. Experimental implementation of our proposed circuits would therefore represent a significant advance. Looking at the whole batch of  improved solutions we proposed, we notice how they are often associated with an input state which requires bunched photons. Realizing multiphoton Fock sources that are more efficient than the current state of the art is therefore another natural objective that will be hopefully pursued by the experimental community.

\textbf{Scientific understanding through AI} -- The ideas discussed here are the result of an automated approach. We showed that automated search can be used not only to optimize heralded photonic circuits, but also to uncover simple physical principles behind high-probability state generation, therefore extracting pure scientific understanding that is by all means computer-inspired. Looking ahead, we deem crucial the progressive implementation of AI systems that are not only capable of producing improved results, but also of generalizing them to different scenarios, thereby becoming agents of understanding \cite{krenn2022scientific}.

\textbf{Increased search spaces} -- Advances in AI systems capable of generating scientific understanding must be accompanied by equally reliable and expressive physical simulators, since these simulators define the experimental reality on which automated scientific discovery relies. With respect to this work, it would be interesting to expand the search space by introducing conditional operations (based on mid-circuit measurements) and active non-linear elements \cite{mccusker2009efficient,engelkemeier2020quantum}, and to model experimental imperfections directly into the optimization, instead of simply analyzing their effects on optimized solutions. Finally, it would be noteworthy to simulate and optimize complete experimental quantum tasks, instead of limiting ourselves to the heralding of the required resource states; such a wider approach could have the potential to completely change our perspective and produce unexpected and surprising ideas. 

\prlheading[sec:Acknowledgments]{Acknowledgments}
M.K. and M.A. acknowledge support by the European Research Council (ERC) under the European Union’s Horizon Europe research and innovation programme (ERC-2024-STG, 101165179, ArtDisQ) and from the German Research Foundation DFG (EXC 2064/1, Project 390727645). X.G. acknowledges support from the Alexander von Humboldt Foundation and the Deutsche Forschungsgemeinschaft (DFG, German Research Foundation) – Project-ID 398816777 – SFB 1375. P.G.K. and C.P.L. acknowledge that this material is based upon work supported by the National Science Foundation under Award No. 25-15092 and  by the National Aeronautics and Space Administration under STTR Contract No. 80NSSC25CA031.

\clearpage 
\bibliographystyle{apsrev4-2} 
\bibliography{biblio}

\clearpage
\onecolumngrid

\begin{center}
{\large\bfseries Supplemental Material for:\\[0.3em]
Automated discovery of high-probability heralded schemes for path-entangled states}
\end{center}

\vspace{1em}

\supsection{Optimization procedure}
\label{app:optimization}

We consider an $M$-mode passive linear-optical circuit decomposed into phase shifters and beam splitters. A phase shifter on mode $i$ acts as
\begin{equation*}
    P_i(\varphi)\ket{n_i}
    =
    e^{i n_i \varphi}\ket{n_i}.
\end{equation*}
For a beam splitter acting on modes $i,j$, we use the convention
\begin{equation*}
    \hat a_i^\dagger
    \longmapsto
    t\,\hat a_i^\dagger
    +
    r\,\hat a_j^\dagger,
    \qquad
    \hat a_j^\dagger
    \longmapsto
    -r^*\,\hat a_i^\dagger
    +
    t\,\hat a_j^\dagger,
\end{equation*}
where
\begin{equation*}
    t=\cos\theta,
    \qquad
    r=e^{i\phi}\sin\theta .
\end{equation*}

\noindent In the two-mode Fock basis, the same beam splitter is evaluated exactly as
\begin{equation*}
    B_{ij}(\theta,\phi)\ket{m,n}
    =
    \sum_{u=0}^{m+n}
    A_{m,n}^{(u)}(\theta,\phi)
    \ket{u,m+n-u},
\end{equation*}
with
\begin{equation*}
    A_{m,n}^{(u)}
    =
    \sqrt{
    \frac{u!\,(m+n-u)!}{m!\,n!}
    }
    \sum_{p=p_{\min}}^{p_{\max}}
    \binom{m}{p}
    \binom{n}{u-p}
    t^{n+2p-u}
    r^{m-p}
    (-r^*)^{u-p},
\end{equation*}
where
\begin{equation*}
    p_{\min}=\max(0,u-n),
    \qquad
    p_{\max}=\min(m,u).
\end{equation*}

\noindent The full interferometer is obtained by composing these elementary maps according to a Reck or Clements mesh. In addition, the total photon number is conserved. Therefore, for a fixed input product Fock state
\begin{equation*}
    \ket{\mathbf n_{\mathrm{in}}}
    =
    \ket{n_1,n_2,\ldots,n_M},
    \qquad
    N=\sum_{i=1}^{M} n_i ,
\end{equation*}
the simulation is restricted to the fixed-$N$ Fock sector
\begin{equation*}
    \mathcal H_{N,M}
    =
    \mathrm{span}
    \left\{
    \ket{\mathbf n}:
    \sum_{i=1}^{M} n_i=N
    \right\}.
\end{equation*}

\noindent Let $\mathcal T$ denote the set of target modes and $\mathcal D$ the set of detected modes, with $\mathcal T\cap\mathcal D=\varnothing$. The desired target state is written as
\begin{equation*}
    \ket{\psi_\star}
    =
    \sum_{\mathbf a}
    \alpha_{\mathbf a}\ket{\mathbf a}_{\mathcal T},
    \qquad
    \sum_{\mathbf a}|\alpha_{\mathbf a}|^2=1 .
\end{equation*}
After the interferometer $U(\boldsymbol\xi)$, where $\boldsymbol\xi$ denotes all beam-splitter and phase-shifter parameters, the global output state is
\begin{equation*}
    \ket{\Psi(\boldsymbol\xi)}
    =
    U(\boldsymbol\xi)\ket{\mathbf n_{\mathrm{in}}}.
\end{equation*}

\noindent For a photon-number-resolving detector pattern $\mathbf h$ on the detected modes, we define the corresponding unnormalized conditional state on the target modes as
\begin{equation*}
    \ket{\psi_{\mathbf h}(\boldsymbol\xi)}
    =
    {}_{\mathcal D}\!\bra{\mathbf h}
    \Psi(\boldsymbol\xi)\rangle .
\end{equation*}
Its heralding probability is
\begin{equation*}
    p_{\mathbf h}(\boldsymbol\xi)
    =
    \norm{\ket{\psi_{\mathbf h}(\boldsymbol\xi)}}^2 ,
\end{equation*}
while its overlap with the desired target state is
\begin{equation*}
    q_{\mathbf h}(\boldsymbol\xi)
    =
    \left|
    \braket{\psi_\star}{\psi_{\mathbf h}(\boldsymbol\xi)}
    \right|^2 .
\end{equation*}
For a single accepted heralding pattern, the conditional fidelity would therefore be
\begin{equation*}
    F_{\mathbf h}(\boldsymbol\xi)
    =
    \frac{
    q_{\mathbf h}(\boldsymbol\xi)
    }{
    p_{\mathbf h}(\boldsymbol\xi)
    } .
\end{equation*}
\noindent In the optimization, however, we do not choose a single heralding pattern in advance. Instead, we define a finite bank of possible detector outcomes,
\begin{equation*}
    \mathcal H
    =
    \{
    \mathbf h_1,\mathbf h_2,\ldots,\mathbf h_W
    \},
\end{equation*}
and assign to each pattern a trainable real logit $u_{\mathbf h}$. The corresponding soft acceptance weight is
\begin{equation*}
    w_{\mathbf h}
    =
    \sigma(u_{\mathbf h})
    =
    \frac{1}{1+e^{-u_{\mathbf h}}}.
\end{equation*}
The total heralding probability used during training is then
\begin{equation*}
    \psucc(\boldsymbol\xi,\mathbf u)
    =
    \sum_{\mathbf h\in\mathcal H}
    w_{\mathbf h}\,
    p_{\mathbf h}(\boldsymbol\xi),
\end{equation*}
and the corresponding fidelity is
\begin{equation*}
    F(\boldsymbol\xi,\mathbf u)
    =
    \frac{
    \sum_{\mathbf h\in\mathcal H}
    w_{\mathbf h}\,
    q_{\mathbf h}(\boldsymbol\xi)
    }{
    \psucc(\boldsymbol\xi,\mathbf u)
    }.
\end{equation*}
Thus the discrete problem of selecting an accepted set of detector outcomes is embedded into a continuous problem through the weights $w_{\mathbf h}$.

The loss function used in the automated discovery is
\begin{equation}
    \mathcal L(\boldsymbol\xi,\mathbf u)
    =
    1
    -
    F(\boldsymbol\xi,\mathbf u)
    +
    \lambda_p
    \frac{1}{
    \psucc(\boldsymbol\xi,\mathbf u)
    },
    \label{eq:fidfirst_loss}
\end{equation}
with $\lambda_p>0$. The first term enforces fidelity with the target state, while the second term penalizes solutions whose high fidelity is obtained only at vanishing heralding probability.

This construction converts the original mixed discrete--continuous search into a differentiable optimization problem. The beam-splitter angles, beam-splitter phases, phase-shifter phases, and heralding logits are all continuous trainable variables
\begin{equation*}
    \left\{
    \theta_\ell,\phi_\ell,\varphi_\ell,u_{\mathbf h}
    \right\}.
\end{equation*}
\noindent This implementation is well suited to JAX because the computational graph is fixed once $M$, the input photon number, the target modes, and the detector-pattern bank are specified. The Fock basis, beam-splitter transition tables, detector-pattern maps, and heralding groups are precomputed as static arrays, while only the optical parameters and heralding logits remain trainable. The forward pass is then expressed as a sequence of array operations over fixed-shape tensors: propagation through the interferometer, grouping of amplitudes by detector outcome, evaluation of $p_{\mathbf h}$ and $q_{\mathbf h}$, and finally evaluation of Eq.~\eqref{eq:fidfirst_loss}. This allows the whole loss and its gradients to be compiled with XLA (Accelerated Linear Algebra) through JAX's just-in-time compilation and executed efficiently on GPUs. Gradients with respect to all continuous variables are obtained by automatic differentiation, and the resulting objective is optimized with the Adam optimizer \cite{kingma2014adam} from many random initializations.

\supsection{Modular comb family}
\label{app:modular_comb_family}

\supsubsection{Explicit derivation}
\label{app:explicit derivation}
\noindent We explicitly derive the general modular-comb construction of the discovered family.  The target photon number is written as
\begin{equation*}
    N
    =
    \sum_{\ell=1}^{L}
    r_\ell m_\ell .
\end{equation*}
The input consists of $L$ bunched packets and of the ancillary photons used by the Fock filters,
\begin{equation*}
    \ket{\psi_{\rm in}}
    =
    \bigotimes_{\ell=1}^{L}
    \ket{m_\ell}^{\otimes r_\ell}
    \otimes
    \ket{1,1}^{\otimes |B|}.
\end{equation*}
The first factor generates the modular comb on the two target modes, while each ancillary pair $\ket{1,1}$ implements one symmetric Fock filter on the two rails. The set $B$ of filters is defined below. We require $r_\ell\geq 2$, so that each packet can coherently populate both target rails.

Let the input creation operators of the $\ell$-th packet be
\begin{equation*}
    \hat u_{\ell,0}^\dagger,
    \hat u_{\ell,1}^\dagger,
    \ldots,
    \hat u_{\ell,r_\ell-1}^\dagger ,
\end{equation*}
and the output modes denoted by
\begin{equation*}
    \hat o_{\ell,0}^\dagger,
    \hat o_{\ell,1}^\dagger,
    \hat h_{\ell,2}^\dagger,
    \ldots,
    \hat h_{\ell,r_\ell-1}^\dagger ,
\end{equation*}
where $\hat o_{\ell,0}^\dagger$ and $\hat o_{\ell,1}^\dagger$ are the two packet rails that will later be coherently collected into the target modes, while $\hat h_{\ell,k}^\dagger$ will be similarly collected into heralding modes. The multiport transformation is of Fourier type:
\begin{equation*}
    \hat u_{\ell,q}^\dagger
    \longrightarrow
    \frac{1}{\sqrt{r_\ell}}
    \left(
        \hat o_{\ell,0}^\dagger
        +
        \zeta_{r_\ell}^{q}\hat o_{\ell,1}^\dagger
        +
        \sum_{k=2}^{r_\ell-1}
        \zeta_{r_\ell}^{qk}\hat h_{\ell,k}^\dagger
    \right),
    \qquad
    \zeta_{r_\ell}
    =
    e^{2\pi i/r_\ell}.
\end{equation*}

\noindent The two packet rails are then combined, for each target rail separately, into the final target modes $\hat a_0^\dagger$ and $\hat a_1^\dagger$. We choose real weights $\lambda_\ell\geq 0$ such that
\begin{equation*}
    \sum_{\ell=1}^{L}
    \lambda_\ell
    =
    1.
\end{equation*}
For $\mu=0,1$, the collection unitary is chosen so that
\begin{equation*}
    \hat o_{\ell,\mu}^\dagger
    \longrightarrow
    \sqrt{\lambda_\ell}\,\hat a_\mu^\dagger
    +
    \sum_{s=1}^{L-1}
    w_{\mu,s\ell}\,
    \hat g_{\mu,s}^\dagger .
\end{equation*}
The modes $\hat g_{\mu,s}^\dagger$ are additional heralding modes. The coefficients $w_{\mu,s\ell}$ complete the normalized row
\begin{equation*}
    \left(
        \sqrt{\lambda_1},
        \sqrt{\lambda_2},
        \ldots,
        \sqrt{\lambda_L}
    \right)
\end{equation*}
to an $L$-mode passive unitary. Combining the multiports and the collection unitaries, the full state before the vacuum heralding is
\begin{equation*}
    \ket{\Psi_{\rm pre}}
    =
    \prod_{\ell=1}^{L}
    \prod_{q=0}^{r_\ell-1}
    \frac{1}{\sqrt{m_\ell!}}
    \Bigg[
        \frac{1}{\sqrt{r_\ell}}
        \Bigg(
            \sqrt{\lambda_\ell}\,\hat a_0^\dagger
            +
            \sum_{s=1}^{L-1}
            w_{0,s\ell}\hat g_{0,s}^\dagger +
            \zeta_{r_\ell}^{q}
            \left(
                \sqrt{\lambda_\ell}\,\hat a_1^\dagger
                +
                \sum_{s=1}^{L-1}
                w_{1,s\ell}\hat g_{1,s}^\dagger
            \right)
            +
            \sum_{k=2}^{r_\ell-1}
            \zeta_{r_\ell}^{qk}
            \hat h_{\ell,k}^\dagger
        \Bigg)
    \Bigg]^{m_\ell}
    \ket{0}.
\end{equation*}
The vacuum heralding condition projects all modes
\begin{equation*}
    \hat h_{\ell,k}^\dagger
    \quad
    (k=2,\ldots,r_\ell-1),
    \qquad
    \hat g_{\mu,s}^\dagger
    \quad
    (\mu=0,1,\;s=1,\ldots,L-1)
\end{equation*}
onto vacuum. The resulting unnormalized two-mode state is therefore
\begin{equation*}
    \ket{\psi_{\rm tar}(\boldsymbol{\lambda})}
    =
    C(\boldsymbol{\lambda})
    \prod_{\ell=1}^{L}
    \prod_{q=0}^{r_\ell-1}
    \left(
        \hat a_0^\dagger
        +
        \zeta_{r_\ell}^{q}
        \hat a_1^\dagger
    \right)^{m_\ell}
    \ket{0},
\end{equation*}
with
\begin{equation*}
    C(\boldsymbol{\lambda})
    =
    \prod_{\ell=1}^{L}
    \frac{
        \left(\lambda_\ell/r_\ell\right)^{r_\ell m_\ell/2}
    }{
        \left(m_\ell!\right)^{r_\ell/2}
    }.
\end{equation*}
Using the identity
\begin{equation*}
    \prod_{q=0}^{r-1}
    \left(
        x+\zeta_r^q y
    \right)
    =
    x^r+(-1)^{r+1}y^r ,
\end{equation*}
we obtain
\begin{equation*}
    \ket{\psi_{\rm tar}(\boldsymbol{\lambda})}
    =
    C(\boldsymbol{\lambda})
    \prod_{\ell=1}^{L}
    \left[
        \left(\hat a_0^\dagger\right)^{r_\ell}
        +
        (-1)^{r_\ell+1}
        \left(\hat a_1^\dagger\right)^{r_\ell}
    \right]^{m_\ell}
    \ket{0}.
\end{equation*}
Expanding each packet, every term contributes to a two-mode sector of the form
\begin{equation*}
    \ket{N-q(\boldsymbol j),q(\boldsymbol j)},
    \qquad
    q(\boldsymbol j)
    =
    \sum_{\ell=1}^{L}
    r_\ell j_\ell ,
    \qquad
    j_\ell=0,\ldots,m_\ell .
\end{equation*}
The set of photon numbers that can arise before contributions to the same sector are coherently combined is
\begin{equation*}
    \widetilde Q
    =
    \left\{
    \sum_{\ell=1}^{L}
    r_\ell j_\ell
    :
    j_\ell=0,\ldots,m_\ell
    \right\}.
\end{equation*}
To determine the actual photon-number support, define the polynomial
\begin{equation*}
    P(z)
    =
    \prod_{\ell=1}^{L}
    \left[
        1+
        (-1)^{r_\ell+1}z^{r_\ell}
    \right]^{m_\ell}
    =
    \sum_{q=0}^{N}
    A_q z^q .
\end{equation*}
The conditional state can then be written as
\begin{equation*}
    \ket{\psi_{\rm tar}(\boldsymbol{\lambda})}
    =
    C(\boldsymbol{\lambda})
    \sum_{q=0}^{N}
    A_q
    \left(\hat a_0^\dagger\right)^{N-q}
    \left(\hat a_1^\dagger\right)^q
    \ket{0}.
\end{equation*}
The photon-number support is therefore the modular comb
\begin{equation*}
    Q
    =
    \left\{
        q\in\widetilde Q
        :
        A_q\neq 0
    \right\}.
\end{equation*}
Since
\begin{equation*}
    A_0=1,
    \qquad
    A_N
    =
    (-1)^{N+\sum_{\ell=1}^{L}m_\ell},
\end{equation*}
the two desired edge sectors $q=0$ and $q=N$ are always present. All other populated sectors are removed by single-photon Fock filters. Since the sectors $q$ and $N-q$ are removed by the same symmetric filter, the distinct filter set is
\begin{equation*}
    B
    =
    \left\{
    \min(q,N-q)
    :
    q\in Q,\;
    q\neq 0,N
    \right\}.
\end{equation*}
Different tuples $\boldsymbol j$ may give the same value of $q$. Their contributions are coherently combined in the coefficient $A_q$ and may cancel exactly. This does not affect the construction, because the filters act on photon-number sectors rather than on the individual paths that contribute to their amplitudes.

\noindent We now derive the action of one filter. Consider one target mode $\hat a_\mu^\dagger$ and one ancillary mode $\hat c^\dagger$, mixed on a beam splitter with real amplitudes
\begin{equation*}
    t=\cos\theta,
    \qquad
    s=\sin\theta.
\end{equation*}
Using the convention we introduced in the previous section we get
\begin{equation*}
    \hat a_\mu^\dagger
    \longrightarrow
    t\hat a_\mu^\dagger+s\hat c^\dagger,
    \qquad
    \hat c^\dagger
    \longrightarrow
    -s\hat a_\mu^\dagger+t\hat c^\dagger .
\end{equation*}
Starting from $k$ photons in the target mode and one photon in the ancillary mode, and heralding again on one photon in the ancillary output, the target component is multiplied by
\begin{equation*}
    f(k)
    =
    t^{k-1}
    \left(
        t^2-ks^2
    \right).
\end{equation*}
To remove a sector in which one rail contains $b$ photons, we choose
\begin{equation*}
    t_b^2
    =
    \frac{b}{b+1},
    \qquad
    s_b^2
    =
    \frac{1}{b+1},
\end{equation*}
then impose $f(b)=0$. Applying the same filter to both rails removes both $\ket{N-b,b}$ and $\ket{b,N-b}$.

On an edge component, the two target rails contain $N$ and $0$ photons. A filter tuned to $b$ therefore contributes the edge amplitude
\begin{equation*}
    f(N)f(0)
    =
    -t_b^N
    \frac{N-b}{b+1}.
\end{equation*}
Thus the edge survival probability of one symmetric filter is
\begin{equation*}
    \left|f(N)f(0)\right|^2
    =
    \left(
        \frac{b}{b+1}
    \right)^N
    \left(
        \frac{N-b}{b+1}
    \right)^2 .
\end{equation*}

\noindent The unfiltered probability weight of the two edge sectors is
\begin{equation*}
    p_{\rm edge}(\boldsymbol{\lambda})
    =
    2N!
    \prod_{\ell=1}^{L}
    \frac{
        \left(\lambda_\ell/r_\ell\right)^{r_\ell m_\ell}
    }{
        \left(m_\ell!\right)^{r_\ell}
    }.
\end{equation*}
Maximizing this expression over $\lambda_\ell$, under the constraint $\sum_\ell\lambda_\ell=1$, gives
\begin{equation*}
    \lambda_\ell
    =
    \frac{r_\ell m_\ell}{N}.
\end{equation*}
With this choice,
\begin{equation*}
    p_{\rm edge}^{(L)}
    =
    2
    \frac{N!}{N^N}
    \prod_{\ell=1}^{L}
    \left(
        \frac{m_\ell^{m_\ell}}{m_\ell!}
    \right)^{r_\ell}.
\end{equation*}
Multiplying by the survival probability of every distinct filter $b\in B$, we obtain
\begin{equation}
    \psucc^{(L)}
    =
    2
    \frac{N!}{N^N}
    \prod_{\ell=1}^{L}
    \left(
        \frac{m_\ell^{m_\ell}}{m_\ell!}
    \right)^{r_\ell}
    \prod_{b\in B}
    \left(
        \frac{b}{b+1}
    \right)^N
    \left(
        \frac{N-b}{b+1}
    \right)^2 .
    \label{eq:S2_general_comb_probability}
\end{equation}
This is Eq. \eqref{eq:fourier_comb_family_probability_text} in the main text.

\supsubsection{Asymptotic ratios}
\label{app:asymptotic_two_modes}
\noindent We derive the asymptotic behavior used in the main text. The optimized modular-comb family is obtained by maximizing over the allowed decompositions of $N$ and over the corresponding filter sets. Therefore, any explicit branch of the family gives a lower bound on the optimized envelope. We show that regular nontrivial branches already give an exponential improvement over Pryde--White and a super-exponential improvement over Zou--Pahlke--Mathis.

\noindent We consider a fixed-shape branch of the construction, in which
$L$ and all $m_\ell$ are fixed, while the integers $r_\ell$ satisfy
\begin{equation*}
    N
    =
    \sum_{\ell=1}^{L}
    r_\ell m_\ell
\end{equation*}
and scale as
\begin{equation*}
    r_\ell
    =
    \rho_\ell N
    +
    O(1),
    \qquad
    \rho_\ell>0,
    \qquad
    \sum_{\ell=1}^{L}
    \rho_\ell m_\ell
    =
    1.
\end{equation*} We also assume that the branch is nontrivial, namely that at least one packet has $m_\ell>1$. If all $m_\ell=1$, the construction reduces to a single-photon multiport branch and does not produce the new enhancement.

The Pryde--White probability is
\begin{equation*}
p_{\rm PW}
=
2\frac{N!}{N^N}.
\end{equation*}
Dividing Eq.~\eqref{eq:S2_general_comb_probability} by $p_{\rm PW}$ gives
\begin{equation*}
\frac{\psucc^{(L)}}{p_{\rm PW}}
=
\prod_{\ell=1}^{L}
\left(
\frac{m_\ell^{m_\ell}}{m_\ell!}
\right)^{r_\ell}
\prod_{b\in B}
\left(
\frac{b}{b+1}
\right)^N
\left(
\frac{N-b}{b+1}
\right)^2 .
\end{equation*}
The first product gives the packet enhancement. Taking the base-ten logarithm,
\begin{equation*}
\sum_{\ell=1}^{L}
r_\ell
\log_{10}
\left(
\frac{m_\ell^{m_\ell}}{m_\ell!}
\right)
=
N
\sum_{\ell=1}^{L}
\rho_\ell
\log_{10}
\left(
\frac{m_\ell^{m_\ell}}{m_\ell!}
\right)
+
O(1).
\end{equation*}
The coefficient is strictly positive for every nontrivial branch, because
\begin{equation*}
\frac{m_\ell^{m_\ell}}{m_\ell!}>1
\qquad
(m_\ell>1).
\end{equation*}

\noindent We now check the filter contribution. Since $L$ and all $m_\ell$ are fixed, the number of candidate sectors is bounded by
\begin{equation*}
    \prod_{\ell=1}^{L}(m_\ell+1),
\end{equation*}
and hence the number of populated interior sectors and distinct filters is bounded independently of $N$. For every filter value present along the branch,
\begin{equation*}
    b=\beta_bN+O(1),
    \qquad
    0<\beta_b\leq\frac{1}{2},
\end{equation*}
where the upper bound follows from the symmetric definition $b=\min(q,N-q)$. Therefore,
\begin{equation*}
N\log_{10}
\left(
\frac{b}{b+1}
\right)
=
-\frac{1}{\beta_b\ln 10}
+
O!\left(\frac1N\right),
\end{equation*}
and
\begin{equation*}
2\log_{10}
\left(
\frac{N-b}{b+1}
\right)
=
2\log_{10}
\left(
\frac{1-\beta_b}{\beta_b}
\right)
+
O!\left(\frac1N\right).
\end{equation*}
Thus each filter contributes a constant factor asymptotically. Since the number of filters is fixed on a fixed-shape branch, the full filter product contributes only $O(1)$ to the logarithm of the ratio. Hence
\begin{equation*}
\log_{10}
\left(
\frac{\psucc^{(L)}}{p_{\rm PW}}
\right)
=
c_{\rm PW}N+O(1),
\end{equation*}
with
\begin{equation*}
c_{\rm PW}
=
\sum_{\ell=1}^{L}
\rho_\ell
\log_{10}
\left(
\frac{m_\ell^{m_\ell}}{m_\ell!}
\right)
>
0 .
\end{equation*}
Equivalently,
\begin{equation*}
\frac{\psucc^{(L)}}{p_{\rm PW}}
=
10^{c_{\rm PW}N+O(1)} .
\end{equation*}
Therefore every nontrivial regular branch improves exponentially over Pryde--White. Since the optimized envelope is at least as large as any explicit branch, the optimized modular-comb family inherits an at least exponential improvement.

\vspace{0.5cm}
\noindent We now compare with the even-$N$ family of Zou, Pahlke, and Mathis. This family is obtained as the limiting case $L=1$, $r=2$, and $m=N/2$. From Eq.~\eqref{eq:S2_general_comb_probability}, its success probability is
\begin{equation*}
p_{\rm ZPM}
=
2
\frac{N!}{2^N\left[\left(\frac N2\right)!\right]^2}
\prod_{j=1}^{\lfloor N/4\rfloor}
\left(
\frac{2j}{2j+1}
\right)^N
\left(
\frac{N-2j}{2j+1}
\right)^2 .
\end{equation*}
The leading asymptotic behavior is controlled by the product
\begin{equation*}
N
\sum_{j=1}^{\lfloor N/4\rfloor}
\log
\left(
\frac{2j}{2j+1}
\right).
\end{equation*}
Using
\begin{equation*}
\prod_{j=1}^{M}
\frac{2j}{2j+1}
=
\frac{4^M(M!)^2}{(2M+1)!},
\end{equation*}
and Stirling's formula, we obtain
\begin{equation*}
\sum_{j=1}^{M}
\log
\left(
\frac{2j}{2j+1}
\right)
=
-\frac12\log M+O(1).
\end{equation*}
With $M=\lfloor N/4\rfloor$, this gives
\begin{equation*}
N
\sum_{j=1}^{\lfloor N/4\rfloor}
\log
\left(
\frac{2j}{2j+1}
\right)
=
-\frac N2\log N+O(N).
\end{equation*}

\noindent The remaining filter contribution is
\begin{equation*}
2
\sum_{j=1}^{\lfloor N/4\rfloor}
\log
\left(
\frac{N-2j}{2j+1}
\right).
\end{equation*}
This term is only $O(N)$. To see this explicitly, first take $N=4M$. Then
\begin{equation*}
\prod_{j=1}^{M}
\frac{N-2j}{2j+1}
=
\prod_{j=1}^{M}
\frac{4M-2j}{2j+1}
=
\frac{4^M}{2(2M+1)} .
\end{equation*}
Therefore
\begin{equation*}
\sum_{j=1}^{M}
\log
\left(
\frac{4M-2j}{2j+1}
\right)
=
M\log 4
-
\log!\left[2(2M+1)\right]
=
O(N).
\end{equation*}
For the other even case, $N=4M+2$, the same product gives
\begin{equation*}
\prod_{j=1}^{M}
\frac{N-2j}{2j+1}
=
\prod_{j=1}^{M}
\frac{4M+2-2j}{2j+1}
=
\frac{4^M}{2M+1},
\end{equation*}
and again the logarithm is $O(N)$. Thus the second part of the filter product contributes only $O(N)$ to $\log p_{\rm ZPM}$.

The factorial prefactor
\begin{equation*}
2
\frac{N!}{2^N\left[\left(\frac N2\right)!\right]^2}
\end{equation*}
contributes only $O(\log N)$ to the logarithm. Thus
\begin{equation*}
\log p_{\rm ZPM}
=
-\frac N2\log N+O(N).
\end{equation*}

\noindent By contrast, every regular nontrivial modular-comb branch derived above has
\begin{equation*}
\log \psucc^{(L)}
=
O(N).
\end{equation*}
Therefore,
\begin{equation*}
\log_{10}
\left(
\frac{\psucc^{(L)}}{p_{\rm ZPM}}
\right)
=
\frac N2\log_{10}N+O(N),
\end{equation*}
or
\begin{equation*}
\frac{\psucc^{(L)}}{p_{\rm ZPM}}
=
10^{\frac N2\log_{10}N+O(N)} .
\end{equation*}
This is a super-exponential improvement in $N$. Since the optimized modular-comb family maximizes over the available decompositions and filters, this gives a lower-bound scaling for the optimized envelope.

\noindent Finally, we specify the finite-$N$ curves used in Fig.~\ref{fig:improvement}. For each photon number $N_i$, we compute the exact optimized ratios
\begin{equation*}
R_{\rm PW}(N_i)
=
\frac{\psucc^{\rm opt}(N_i)}{p_{\rm PW}(N_i)},
\qquad
R_{\rm ZPM}(N_i)
=
\frac{\psucc^{\rm opt}(N_i)}{p_{\rm ZPM}(N_i)} .
\end{equation*}
The Pryde--White ratio is fitted to the exponential form
\begin{equation*}
\log_{10} R_{\rm PW}^{\rm fit}(N)
=
a_{\rm PW}N+b_{\rm PW},
\end{equation*}
or equivalently
\begin{equation*}
R_{\rm PW}^{\rm fit}(N)
=
10^{a_{\rm PW}N+b_{\rm PW}} .
\end{equation*}
The Zou--Pahlke--Mathis ratio is fitted to the super-exponential form
\begin{equation*}
\log_{10} R_{\rm ZPM}^{\rm fit}(N)
=
\frac N2\log_{10}N
+
a_{\rm ZPM}N
+
b_{\rm ZPM},
\end{equation*}
or equivalently
\begin{equation*}
R_{\rm ZPM}^{\rm fit}(N)
=
10^{
\frac N2\log_{10}N
+
a_{\rm ZPM}N
+
b_{\rm ZPM}
} .
\end{equation*}
The constants $a_{\rm PW}$, $b_{\rm PW}$, $a_{\rm ZPM}$, and $b_{\rm ZPM}$ represent the $O(N)$ and $O(1)$ terms left unspecified by the asymptotic derivation. The fitted coefficients used for the curves in Fig.~\ref{fig:improvement} are reported in Table~\ref{tab:S2_two_mode_fit_parameters}. 

\begin{table}[t]
\centering
\small
\caption{Finite-$N$ fit parameters used for the solid curves in Fig.~\ref{fig:improvement}. The fitted curves are $R_{\rm PW}^{\rm fit}(N)=10^{a_{\rm PW}N+b_{\rm PW}}$ and $R_{\rm ZPM}^{\rm fit}(N)=10^{\frac N2\log_{10}N+a_{\rm ZPM}N+b_{\rm ZPM}}$.}
\label{tab:S2_two_mode_fit_parameters}
\begin{tabular}{c c c}
\toprule
Ratio & $a$ & $b$ \\
\midrule
$\psucc/p_{\rm PW}$ & $0.272830$ & $-2.770868$ \\
$\psucc/p_{\rm ZPM}$ & $-0.670268$ & $1.566354$ \\
\bottomrule
\end{tabular}
\end{table}

\supsubsection{\texorpdfstring{$\NOON_9$}{NOON-9} example}
\label{app:noon9}
\noindent We now apply the modular-comb construction explicitly to the $\NOON_9$ case. We list all inequivalent packet decompositions of
\begin{equation*}
9
=
\sum_{\ell=1}^{L}
r_\ell m_\ell ,
\qquad
r_\ell\geq 2,
\qquad
m_\ell\geq 1,
\end{equation*}
where decompositions related only by a permutation of the packets are identified. For each decomposition, the comb support $Q$ is defined as the set
of photon-number sectors with nonzero total coefficient, as in
Sec.~\ref{app:explicit derivation}. The corresponding filter set is
\begin{equation*}
    B
    =
    \left\{
        \min(q,9-q)
        :
        q\in Q,\;
        q\neq 0,9
    \right\}.
\end{equation*}
The corresponding probabilities are obtained directly from Eq.~\eqref{eq:S2_general_comb_probability}, and shown in Table \ref{tab:noon9_decompositions}.

\begin{table}[htbp]
\centering
\small
\caption{All inequivalent packet decompositions of $N=9$ within the modular-comb family, together with the corresponding filter sets and success probabilities. The best choice is the single-packet branch $L=1$, $r_1=3$, $m_1=3$.}
\label{tab:noon9_decompositions}
\begin{tabular}{c c c c}
\toprule
$L$ & $\{(r_\ell,m_\ell)\}_{\ell=1}^{L}$ & $B$ & $\psucc$ \\
\midrule
$1$ & $\{(3,3)\}$ & $\{3\}$ & $2.884\%$ \\
$2$ & $\{(3,1),(3,2)\}$ & $\{3\}$ & $0.253\%$ \\
$1$ & $\{(9,1)\}$ & $\varnothing$ & $0.187\%$ \\
$3$ & $\{(3,1),(3,1),(3,1)\}$ & $\{3\}$ & $0.0316\%$ \\
$2$ & $\{(3,1),(6,1)\}$ & $\{3\}$ & $0.0316\%$ \\
$2$ & $\{(2,1),(7,1)\}$ & $\{2\}$ & $0.0265\%$ \\
$2$ & $\{(4,1),(5,1)\}$ & $\{4\}$ & $0.0251\%$ \\
$2$ & $\{(2,2),(5,1)\}$ & $\{2,4\}$ & $0.0142\%$ \\
$2$ & $\{(2,3),(3,1)\}$ & $\{2,3,4\}$ & $0.0122\%$ \\
$3$ & $\{(2,1),(2,1),(5,1)\}$ & $\{2,4\}$ & $0.00356\%$ \\
$3$ & $\{(2,1),(2,2),(3,1)\}$ & $\{2,3,4\}$ & $0.00241\%$ \\
$4$ & $\{(2,1),(2,1),(2,1),(3,1)\}$ & $\{2,3,4\}$ & $0.000602\%$ \\
$3$ & $\{(2,1),(3,1),(4,1)\}$ & $\{2,3,4\}$ & $0.000602\%$ \\
\bottomrule
\end{tabular}
\end{table}

\noindent The best configuration is therefore
\begin{equation*}
L=1,
\qquad
r_1=3,
\qquad
m_1=3.
\end{equation*}

\noindent We now show explicitly how the circuit in Fig. \ref{fig:noon9_family} produces the target state. The input is
\begin{equation*}
\ket{\psi_{\rm in}}
=
\ket{3,3,3,1,1}.
\end{equation*}
In creation-operator form,
\begin{equation*}
\ket{\psi_{\rm in}}
=
\frac{
(\hat u_0^\dagger)^3
(\hat u_1^\dagger)^3
(\hat u_2^\dagger)^3
\hat c_0^\dagger
\hat c_1^\dagger
}{
\sqrt{(3!)^3}
}
\ket{0},
\end{equation*}
where $\hat c_0^\dagger$ and $\hat c_1^\dagger$ are the ancillary single photons used by the Fock filter. The three bunched modes are first sent through a three-mode Fourier multiport,
\begin{equation*}
\hat u_q^\dagger
\longrightarrow
\frac{1}{\sqrt 3}
(
\hat a^\dagger
+
\omega^q \hat b^\dagger
+
\omega^{2q}\hat h^\dagger
),
\qquad
\omega=e^{2\pi i/3},
\qquad
q=0,1,2.
\end{equation*}
Projecting the heralding mode $\hat h$ onto vacuum gives the unnormalized two-mode state
\begin{equation*}
\ket{\psi_{\rm comb}}
=
\frac{1}{3^{9/2}\sqrt{(3!)^3}}
\prod_{q=0}^{2}
(
\hat a^\dagger
+
\omega^q \hat b^\dagger
)^3
\ket{0}.
\end{equation*}
Using
\begin{equation*}
\prod_{q=0}^{2}
(
x+\omega^q y
)
=
x^3+y^3,
\end{equation*}
we obtain
\begin{equation*}
\ket{\psi_{\rm comb}}
=
C
[
(\hat a^\dagger)^3
+
(\hat b^\dagger)^3
]^3
\ket{0},
\qquad
C=
\frac{1}{3^{9/2}\sqrt{(3!)^3}} .
\end{equation*}
Expanding,
\begin{equation*}
\ket{\psi_{\rm comb}}
=
C
[
(\hat a^\dagger)^9
+
3(\hat a^\dagger)^6
(\hat b^\dagger)^3
+
3(\hat a^\dagger)^3
(\hat b^\dagger)^6
+
(\hat b^\dagger)^9
]
\ket{0}.
\end{equation*}
Equivalently, in normalized Fock states,
\begin{equation*}
\ket{\psi_{\rm comb}}
=
C
[
\sqrt{9!}\ket{9,0}
+
3\sqrt{6!3!}\ket{6,3}
+
3\sqrt{3!6!}\ket{3,6}
+
\sqrt{9!}\ket{0,9}
].
\end{equation*}
Thus the Fourier multiport produces the modular comb
\begin{equation*}
\ket{9,0},
\qquad
\ket{6,3},
\qquad
\ket{3,6},
\qquad
\ket{0,9}.
\end{equation*}

\noindent The filter set is $B={3}$. We therefore apply one symmetric Fock filter tuned to $b=3$ on the two target rails. With our beam-splitter convention, the first target rail and its ancillary mode transform as
\begin{equation*}
\hat a^\dagger
\longrightarrow
t\hat a^\dagger+s\hat c_0^\dagger,
\qquad
\hat c_0^\dagger
\longrightarrow
-s\hat a^\dagger+t\hat c_0^\dagger,
\end{equation*}
and similarly for the second rail,
\begin{equation*}
\hat b^\dagger
\longrightarrow
t\hat b^\dagger+s\hat c_1^\dagger,
\qquad
\hat c_1^\dagger
\longrightarrow
-s\hat b^\dagger+t\hat c_1^\dagger.
\end{equation*}
For a filter tuned to $b=3$,
\begin{equation*}
t^2=\frac34,
\qquad
s^2=\frac14.
\end{equation*}
Heralding again on one photon in the ancillary output multiplies a target component with $k$ photons by
\begin{equation*}
f(k)
=
t^{k-1}
(
t^2-ks^2
).
\end{equation*}
With the above choice,
\begin{equation*}
f(3)
=
t^2(t^2-3s^2)
=
0.
\end{equation*}
Therefore the two middle sectors are removed:
\begin{equation*}
\ket{6,3}
\longrightarrow
f(6)f(3)\ket{6,3}
=
0,
\qquad
\ket{3,6}
\longrightarrow
f(3)f(6)\ket{3,6}
=
0.
\end{equation*}
The two edge sectors survive with the same amplitude,
\begin{equation*}
f(9)f(0)
=
-t^9\frac{9-3}{3+1}
=
-\frac{243\sqrt 3}{1024}.
\end{equation*}
The unnormalized heralded state is therefore
\begin{equation*}
\ket{\psi_{\rm out}}
=
C\sqrt{9!},f(9)f(0)
(
\ket{9,0}
+
\ket{0,9}
).
\end{equation*}
After normalization, this is exactly
\begin{equation*}
\ket{\NOON_9}
=
\frac{
\ket{9,0}
+
\ket{0,9}
}{\sqrt 2}.
\end{equation*}
The success probability is
\begin{equation*}
\psucc
=
2
C^2
9!
|f(9)f(0)|^2
=
\frac{945}{32768},
\end{equation*}
in agreement with Eq. \eqref{eq:S2_general_comb_probability}.

\supsection{Multimode \texorpdfstring{$\NOON$}{NOON} extension}
\label{app:multimode_noon}
\supsubsection{Exact derivation}
\label{app:derivation_multi_mode}
\noindent We now extend the modular-comb construction to multimode $\NOON$ states. We avoid repeating the full two-mode derivation and only describe the ingredients that change in the multimode case. The target state is
\begin{equation*}
    \ket{\NOON_N^{(d)}}
    =
    \frac{1}{\sqrt d}
    \sum_{\mu=0}^{d-1}
    e^{i\phi_\mu}
    \ket{0,\ldots,0,N_\mu,0,\ldots,0},
\end{equation*}
where $\phi_0=0$ fixes the global phase convention and the subscript indicates that the $N$ photons occupy the $\mu$-th target mode.

As before, we write
\begin{equation*}
    N
    =
    \sum_{\ell=1}^{L}
    r_\ell m_\ell .
\end{equation*}
The additional multimode requirement is
\begin{equation*}
    r_\ell\geq d,
\end{equation*}
so that each packet can coherently populate the $d$ target modes.

Let the input creation operators of the $\ell$-th packet be
\begin{equation*}
    \hat u_{\ell,0}^\dagger,
    \hat u_{\ell,1}^\dagger,
    \ldots,
    \hat u_{\ell,r_\ell-1}^\dagger .
\end{equation*}
The packet is sent through an $r_\ell$-mode Fourier multiport. We keep the first $d$ output modes as the packet rails that will be collected into the $d$ target modes, while the remaining outputs are projected onto vacuum. Thus
\begin{equation*}
    \hat u_{\ell,q}^\dagger
    \longrightarrow
    \frac{1}{\sqrt{r_\ell}}
    \biggl[
        \sum_{\mu=0}^{d-1}
        \zeta_{r_\ell}^{q\mu}
        \hat o_{\ell,\mu}^\dagger
        +
        \sum_{k=d}^{r_\ell-1}
        \zeta_{r_\ell}^{qk}
        \hat h_{\ell,k}^\dagger
    \biggr],
    \qquad
    \zeta_{r_\ell}
    =
    e^{2\pi i/r_\ell}.
\end{equation*}
For each target mode $\mu$, the packet rails are collected into the final target mode $\hat a_\mu^\dagger$. As in the two-mode case, we choose weights $\lambda_\ell\geq0$ with
\begin{equation*}
    \sum_{\ell=1}^{L}\lambda_\ell=1,
\end{equation*}
and take the collection unitary to contain the row
\begin{equation*}
    \hat o_{\ell,\mu}^\dagger
    \longrightarrow
    \sqrt{\lambda_\ell}\,\hat a_\mu^\dagger
    +
    \text{heralding modes}.
\end{equation*}
After projecting all heralding modes onto vacuum, the unnormalized target state is
\begin{equation}
    \ket{\psi_{\rm tar}^{(d)}}
    =
    C(\boldsymbol{\lambda})
    \prod_{\ell=1}^{L}
    \prod_{q=0}^{r_\ell-1}
    \biggl[
        \sum_{\mu=0}^{d-1}
        \zeta_{r_\ell}^{q\mu}
        \hat a_\mu^\dagger
    \biggr]^{m_\ell}
    \ket{0},
    \label{eq:S3_target_state_before_filtering}
\end{equation}
with
\begin{equation*}
    C(\boldsymbol{\lambda})
    =
    \prod_{\ell=1}^{L}
    \frac{
        (\lambda_\ell/r_\ell)^{r_\ell m_\ell/2}
    }{
        (m_\ell!)^{r_\ell/2}
    }.
\end{equation*}

\noindent Equation~\eqref{eq:S3_target_state_before_filtering} shows explicitly why the output is a multimode comb. Expanding the product gives only monomials of total degree $N$ in the $d$ target creation operators. Therefore every populated sector has the form
\begin{equation*}
    (\hat a_0^\dagger)^{N-|\mathbf q|}
    (\hat a_1^\dagger)^{q_1}
    \cdots
    (\hat a_{d-1}^\dagger)^{q_{d-1}}
    \ket{0},
    \qquad
    |\mathbf q|
    =
    \sum_{\mu=1}^{d-1}q_\mu .
\end{equation*}
Equivalently, in normalized Fock notation, every populated sector is of the form
\begin{equation}
    \ket{N-|\mathbf q|,q_1,\ldots,q_{d-1}}.
    \label{eq:S3_multimode_comb_sector}
\end{equation}
The allowed vectors $\mathbf q=(q_1,\ldots,q_{d-1})$ are precisely those whose coefficient in Eq.~\eqref{eq:S3_target_state_before_filtering} is nonzero. More explicitly, we may write
\begin{equation*}
    \ket{\psi_{\rm tar}^{(d)}}
    =
    \sum_{\mathbf q\in Q_d}
    A_{\mathbf q}
    \ket{N-|\mathbf q|,q_1,\ldots,q_{d-1}},
\end{equation*}
where $Q_d$ is the multimode comb support generated by the packets. This is the direct multimode analogue of the two-mode support $Q$.

The desired $\NOON$ components are the $d$ edge sectors,
\begin{equation*}
    \ket{N,0,\ldots,0},
    \quad
    \ket{0,N,\ldots,0},
    \quad
    \ldots,
    \quad
    \ket{0,\ldots,0,N}.
\end{equation*}
All other populated sectors are interior sectors and must be removed by Fock filters. For a filter tuned to photon number $b$, the beam splitter is chosen as
\begin{equation*}
    t_b^2
    =
    \frac{b}{b+1},
    \qquad
    s_b^2
    =
    \frac{1}{b+1}.
\end{equation*}
Heralding on one photon in the ancillary output multiplies a target component with $k$ photons by
\begin{equation*}
    f_b(k)
    =
    t_b^{k-1}
    (
        t_b^2-ks_b^2
    ),
\end{equation*}
and therefore removes every component with $k=b$.

In the multimode construction, once a filter value $b$ is selected, the same filter is applied to all $d$ target modes. Let $B_d$ be a set of filter values such that every populated non-edge sector has at least one target mode containing exactly $b$ photons for some $b\in B_d$. Explicitly, for every populated occupation vector
\begin{equation*}
    \mathbf n
    =
    (n_0,n_1,\ldots,n_{d-1})
\end{equation*}
which is not one of the $d$ edge vectors, there must exist a mode $\mu$ and a filter value $b\in B_d$ such that
\begin{equation*}
    n_\mu=b.
\end{equation*}
Then the filters remove all interior sectors and leave only the multimode $\NOON$ state.

On an edge component, one target mode contains $N$ photons and the remaining $d-1$ target modes contain zero photons. A filter tuned to $b$ therefore contributes the edge amplitude
\begin{equation*}
    f_b(N)f_b(0)^{d-1}
    =
    -t_b^{N+d-2}
    \frac{N-b}{b+1}.
\end{equation*}
The corresponding edge survival probability is
\begin{equation}
    F_b^{(d)}(N)
    =
    |f_b(N)f_b(0)^{d-1}|^2
    =
    \biggl(
        \frac{b}{b+1}
    \biggr)^{N+d-2}
    \biggl(
        \frac{N-b}{b+1}
    \biggr)^2 .
    \label{eq:S3_filter_survival}
\end{equation}

\noindent Repeating the edge-weight calculation of Sec.~\ref{app:modular_comb_family}, the unfiltered edge probability is
\begin{equation*}
    p_{\rm edge}^{(d)}
    =
    d
    \frac{N!}{N^N}
    \prod_{\ell=1}^{L}
    \biggl(
        \frac{m_\ell^{m_\ell}}{m_\ell!}
    \biggr)^{r_\ell}.
\end{equation*}
Multiplying by the survival probability of all selected filters gives
\begin{equation}
    \psucc^{(d)}
    =
    d
    \frac{N!}{N^N}
    \prod_{\ell=1}^{L}
    \biggl(
        \frac{m_\ell^{m_\ell}}{m_\ell!}
    \biggr)^{r_\ell}
    \prod_{b\in B_d}
    F_b^{(d)}(N).
    \label{eq:S3_multimode_probability}
\end{equation}
This is the multimode analogue of Eq.~\eqref{eq:S2_general_comb_probability}.

\supsubsection{Asymptotic ratio}
\label{app:asymptotic_multi_modes}

\noindent
We derive the asymptotic behaviour of the multimode modular-comb
family relative to the Zhang--Chan construction. For a fixed pair
$(N,d)$, the optimized modular-comb family is obtained by maximizing
over the allowed packet decompositions of $N$. Therefore, as in the
two-mode case, any explicit fixed-shape branch provides a lower bound
on the optimized envelope.

To compare the two constructions under the same filtering conditions,
we impose the universal filter set used by
Zhang and Chan~\cite{Zhang2018MultimodeNOON},
\begin{equation*}
    B_{\rm univ}
    =
    \{1,2,\ldots,\lfloor N/2\rfloor\}.
\end{equation*}
This set removes every non-edge $d$-mode occupation vector: any such
vector has at least two nonzero entries, and hence at least one entry
between $1$ and $\lfloor N/2\rfloor$.

Under this common filtering prescription, the modular-comb success
probability is
\begin{equation*}
    \psucc^{(d)}
    =
    d
    \frac{N!}{N^N}
    \prod_{\ell=1}^{L}
    \left(
        \frac{m_\ell^{m_\ell}}{m_\ell!}
    \right)^{r_\ell}
    \prod_{b=1}^{\lfloor N/2\rfloor}
    F_b^{(d)}(N),
\end{equation*}
whereas the Zhang--Chan probability is
\begin{equation*}
    p_{\rm ZC}^{(d)}
    =
    d^{1-N}
    \prod_{b=1}^{\lfloor N/2\rfloor}
    F_b^{(d)}(N).
\end{equation*}
The filtering contribution therefore cancels exactly from the ratio,
giving
\begin{equation*}
    R_{\rm ZC}^{(d)}(N)
    =
    \frac{\psucc^{(d)}}{p_{\rm ZC}^{(d)}}
    =
    d^N
    \frac{N!}{N^N}
    \prod_{\ell=1}^{L}
    \left(
        \frac{m_\ell^{m_\ell}}{m_\ell!}
    \right)^{r_\ell}.
\end{equation*}

\noindent
We now consider a fixed-shape branch of the construction, in which
$L$ and all $m_\ell$ are fixed, while the integers $r_\ell$ satisfy
\begin{equation*}
    N
    =
    \sum_{\ell=1}^{L}
    r_\ell m_\ell
\end{equation*}
and scale as
\begin{equation*}
    r_\ell
    =
    \rho_\ell N
    +
    O(1),
    \qquad
    \rho_\ell>0,
    \qquad
    \sum_{\ell=1}^{L}
    \rho_\ell m_\ell
    =
    1.
\end{equation*} Using Stirling's formula, we obtain
\begin{align*}
    \log_{10}
    R_{\rm ZC}^{(d)}(N)
    &=
    N
    \log_{10}
    \left(
        \frac{d}{e}
    \right)
    +
    \frac{1}{2}
    \log_{10}N
    \\
    &\quad+
    \sum_{\ell=1}^{L}
    r_\ell
    \log_{10}
    \left(
        \frac{m_\ell^{m_\ell}}{m_\ell!}
    \right)
    +
    O(1)
    \\
    &=
    c_{\rm ZC}^{(d)}N
    +
    \frac{1}{2}
    \log_{10}N
    +
    O(1),
\end{align*}
where
\begin{equation*}
    c_{\rm ZC}^{(d)}
    =
    \log_{10}
    \left(
        \frac{d}{e}
    \right)
    +
    \sum_{\ell=1}^{L}
    \rho_\ell
    \log_{10}
    \left(
        \frac{m_\ell^{m_\ell}}{m_\ell!}
    \right).
\end{equation*}
The coefficient $c_{\rm ZC}^{(d)}$ depends on the chosen fixed-shape
branch through $\rho_\ell$ and $m_\ell$.

\noindent
For $d\geq3$, this coefficient is strictly positive. Indeed,
$d/e>1$, while
\begin{equation*}
    \frac{m_\ell^{m_\ell}}{m_\ell!}
    \geq
    1
\end{equation*}
for every $m_\ell\geq1$. Therefore every fixed-shape branch satisfies
\begin{equation*}
    R_{\rm ZC}^{(d)}(N)
    =
    N^{1/2}
    10^{
        c_{\rm ZC}^{(d)}N
        +
        O(1)
    }.
\end{equation*}
Since the optimized modular-comb family is at least as successful as
any explicit branch, for any chosen fixed-shape branch its ratio with
respect to the Zhang--Chan construction satisfies
\begin{equation*}
    R_{\rm ZC,opt}^{(d)}(N)
    \geq
    N^{1/2}
    10^{
        c_{\rm ZC}^{(d)}N
        +
        O(1)
    }.
\end{equation*}

\noindent
For Fig.~\ref{fig:multi_mode_improvement}, the points are the exact
optimized ratios, obtained by maximizing over all allowed packet
decompositions for each pair $(N,d)$ under the common universal
filtering prescription. Since the optimal integer decomposition can
change with $N$, the exact envelope need not coincide with a single
fixed-shape branch. The solid curves are finite-$N$ fits to the
lower-bound-inspired form
\begin{equation*}
    \log_{10}
    R_{\rm ZC}^{\rm fit}(N,d)
    =
    \frac{1}{2}
    \log_{10}N
    +
    a_dN
    +
    b_d,
\end{equation*}
or equivalently
\begin{equation*}
    R_{\rm ZC}^{\rm fit}(N,d)
    =
    N^{1/2}
    10^{a_dN+b_d}.
\end{equation*}
The fitted coefficients $a_d$ and $b_d$ are reported in
Table~\ref{tab:S3_multimode_fit_parameters}.

\begin{table}[htbp]
\centering
\small
\caption{
Finite-$N$ fit parameters used for the solid curves in
Fig.~\ref{fig:multi_mode_improvement}. The points are the exact
optimized ratios obtained under the common universal filtering
prescription, whereas the curves are fitted to
$R_{\rm ZC}^{\rm fit}(N,d)=N^{1/2}10^{a_dN+b_d}$.
}
\label{tab:S3_multimode_fit_parameters}
\begin{tabular}{c c c}
\toprule
$d$ & $a_d$ & $b_d$ \\
\midrule
$3$ & $0.408426$ & $-1.479231$ \\
$4$ & $0.507060$ & $-1.823177$ \\
$5$ & $0.593785$ & $-2.182414$ \\
$6$ & $0.648037$ & $-2.453764$ \\
$7$ & $0.705166$ & $-2.687828$ \\
\bottomrule
\end{tabular}
\end{table}

\begin{figure}[htbp]
\centering
\begin{minipage}{0.73\linewidth}
    \centering
    \includegraphics[width=\linewidth]{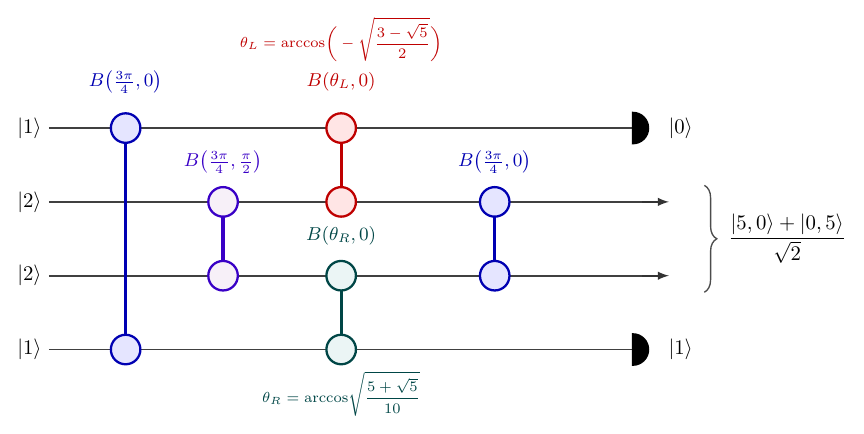}
    \vspace{-0.5em}
\end{minipage}
\hfill
\begin{minipage}{0.73\linewidth}
    \centering
    \includegraphics[width=\linewidth]{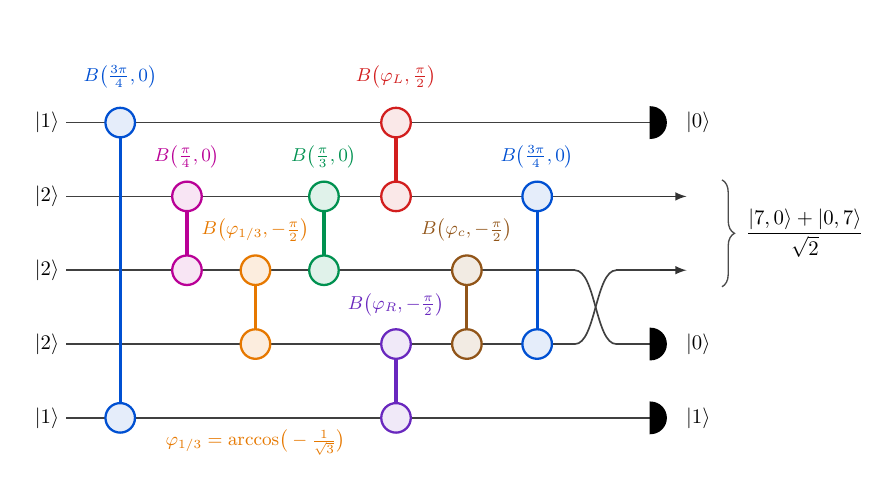}
    \vspace{-0.5em}
\end{minipage}
\hfill
\begin{minipage}{0.73\linewidth}
    \centering
    \includegraphics[width=\linewidth]{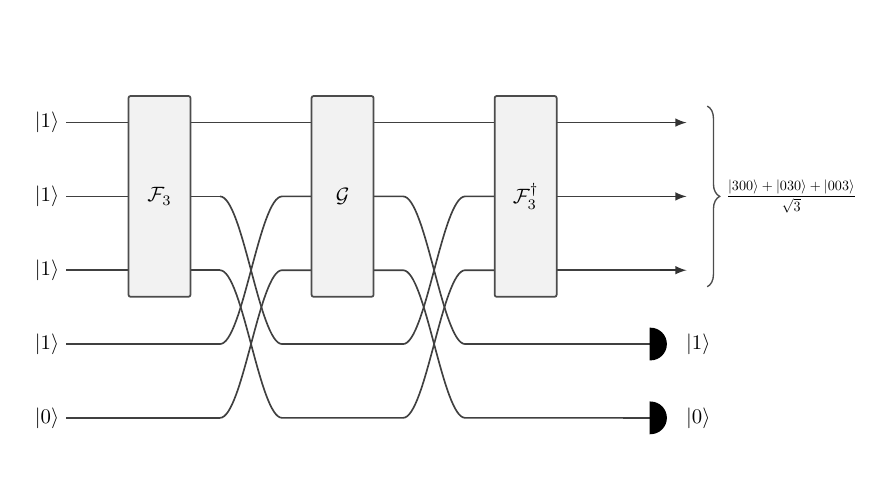}
    \vspace{-0.5em}
\end{minipage}

\caption{ 
Explicit heralded constructions for the two and three mode $\NOON$ states from Table \ref{tab:additional_schemes} that were not analyzed in depth in the main text.
}
\label{fig:additional_schemes}
\end{figure}

\supsection{Additional schemes}
\label{app:additional schemes}
\supsubsection{Two- and three-mode \texorpdfstring{$\NOON$}{NOON} states}
\label{app:additional_noon}
We explicitly show in Fig. \ref{fig:additional_schemes} the three schemes for two and three mode $\NOON$ states reported in Table \ref{tab:additional_schemes} and not discussed in depth in the main text. We now provide additional comments about some of these. For the $\NOON_7$ scheme, not all the parameters are rational numbers, and with
\begin{equation*}
    \begin{pmatrix}
        \varphi_L \\
        \varphi_R \\
        \varphi_c
    \end{pmatrix}
    =
    \begin{pmatrix}
        0.660468132075635\\
        2.655395656573681\\
        0.146643044288461
    \end{pmatrix}
\end{equation*}
we get
\begin{equation*}
    (1- F)
    <
    10^{-15},
    \qquad
    \psucc
    =
    0.0312535870177935,
\end{equation*}
essentially $F=1$ and $\psucc = 3.125\%$ to double numerical precision. To check the robustness of this solution, we can consider some sharp simplifications for the parameters. For instance, taking
\begin{equation*}
    \begin{pmatrix}
        \varphi_L \\
        \varphi_R \\
        \varphi_c
    \end{pmatrix}
    =
    \begin{pmatrix}
        \arccos\bigg({\sqrt{\frac{5}{8}}\bigg)}\\
        \arccos\bigg({-\sqrt{\frac{43}{55}}\bigg)}\\
        \arccos\bigg({\sqrt{\frac{46}{47}}\bigg)}
    \end{pmatrix}
\end{equation*}

yields
\begin{equation*}
    (1-F)
    <
    10^{-6},
    \qquad
    \psucc
    =
    0.03128111099616.
\end{equation*}

\noindent For the \(N=3\), three-mode $\NOON$ state, the central block
\(\mathcal G\) represents a single-photon-assisted heralded
operation acting on one target mode and two ancillary modes.
Conditioning on the ancilla output pattern induces a
photon-number-dependent transformation on the target mode.

We label the three target modes by \(0,1,2\), and the two
ancillary modes by \(h,v\). The input state is
\begin{equation*}
    \ket{\psi_{\rm in}}
    =
    \ket{1,1,1}_{012}
    \otimes
    \ket{1,0}_{hv}.
\end{equation*}
The target state is
\begin{equation*}
    \ket{\NOON_{3,3}}_{012}
    =
    \frac{
        \ket{3,0,0}_{012}
        +
        \ket{0,3,0}_{012}
        +
        \ket{0,0,3}_{012}
    }{
        \sqrt{3}
    }.
\end{equation*}

\noindent The first tritter acts only on the target modes. It gives
\begin{equation*}
\begin{aligned}
    (\mathcal F_3 \otimes I_{hv})
    \ket{\psi_{\rm in}}
    &=
    \left[
        \frac{\sqrt{2}}{3}
        \left(
            \ket{3,0,0}_{012}
            +
            \ket{0,3,0}_{012}
            +
            \ket{0,0,3}_{012}
        \right)
        -
        \frac{1}{\sqrt{3}}
        \ket{1,1,1}_{012}
    \right]
    \otimes
    \ket{1,0}_{hv}.
\end{aligned}
\end{equation*}

\noindent The central block acts nontrivially only on modes \(0,h,v\).
We denote its three-mode unitary by \(G_{0hv}\). Conditioning on the
ancillary output pattern \(\ket{1,0}_{hv}\) induces the effective
Kraus operator
\begin{equation*}
    K_0
    =
    {}_{hv}\!\bra{1,0}\,
    G_{0hv}\,
    \ket{1,0}_{hv}.
\end{equation*}
This operator acts diagonally on the photon number in target mode
\(0\):
\begin{equation*}
    K_0\ket{n}_0
    =
    K(n)\ket{n}_0,
    \qquad
    K(n)
    =
    a^{n-1}
    \left(
        a\lambda
        +
        nbc
    \right).
\end{equation*}
We choose the parameters so that
\begin{equation*}
    K(0)
    =
    K(3)
    =
    \lambda,
    \qquad
    K(1)
    =
    -2\lambda.
\end{equation*}
Equivalently, they obey
\begin{equation*}
    bc
    =
    -(a+2)\lambda,
    \qquad
    2a^3
    +
    6a^2
    +
    1
    =
    0,
\end{equation*}
where we select the root satisfying
\begin{equation*}
    \operatorname{Im}(a)>0.
\end{equation*}

\noindent It remains to determine the physical value of \(\lambda\).
We take
\begin{equation*}
    b
    =
    \sqrt{
        \lambda\lvert a+2\rvert
    },
    \qquad
    c
    =
    -(a+2)
    \sqrt{
        \frac{\lambda}{\lvert a+2\rvert}
    },
\end{equation*}
which ensures \(bc=-(a+2)\lambda\). The matrix
\begin{equation*}
    M
    =
    \begin{pmatrix}
        a & b \\
        c & \lambda
    \end{pmatrix}
\end{equation*}
must be a contraction in order to admit a passive
linear-optical implementation through a vacuum extension
\cite{vanmeter2007general}. This condition gives
\begin{equation*}
    \left(
        4\lvert a+1\rvert^2
        -
        1
    \right)
    \lambda^2
    -
    2\lvert a+2\rvert\lambda
    +
    1
    -
    \lvert a\rvert^2
    =
    0.
\end{equation*}
We choose the smaller positive root,
\begin{equation*}
    \lambda
    =
    \frac{
        \lvert a+2\rvert
        -
        \sqrt{
            \lvert a+2\rvert^2
            -
            \left(
                4\lvert a+1\rvert^2
                -
                1
            \right)
            \left(
                1
                -
                \lvert a\rvert^2
            \right)
        }
    }{
        4\lvert a+1\rvert^2
        -
        1
    }.
\end{equation*}
For the selected root of
\(2a^3+6a^2+1=0\), this gives
\begin{equation*}
    \lambda
    =
    0.271156945703\ldots.
\end{equation*}

\noindent The heralded action of the central block on the target modes
is therefore
\begin{align*}
    (K_0\otimes I_{12})
    \mathcal F_3
    \ket{1,1,1}_{012}
    &=
    \lambda
    \left[
        \frac{\sqrt{2}}{3}
        \left(
            \ket{3,0,0}_{012}
            +
            \ket{0,3,0}_{012}
            +
            \ket{0,0,3}_{012}
        \right)
        +
        \frac{2}{\sqrt{3}}
        \ket{1,1,1}_{012}
    \right]
    \\
    &=
    \sqrt{2}\,\lambda\,
    \mathcal F_3
    \ket{\NOON_{3,3}}_{012}.
\end{align*}
Since the final tritter is \(\mathcal F_3^\dagger\), the complete
heralded transformation is
\begin{equation*}
    \mathcal F_3^\dagger
    (K_0\otimes I_{12})
    \mathcal F_3
    \ket{1,1,1}_{012}
    =
    \sqrt{2}\,\lambda\,
    \ket{\NOON_{3,3}}_{012}.
\end{equation*}
The heralding probability is consequently
\begin{equation*}
    \psucc
    =
    2\lvert\lambda\rvert^2
    =
    0.147052178406\ldots.
\end{equation*}

\supsubsection{\texorpdfstring{$m,m'$}{m,m-prime} states}
\label{app:m_m'}

\begin{figure}[htbp]
    \centering
    \makebox[\linewidth][c]{%
        \hspace*{2cm}%
        \includegraphics[width=0.8\linewidth]{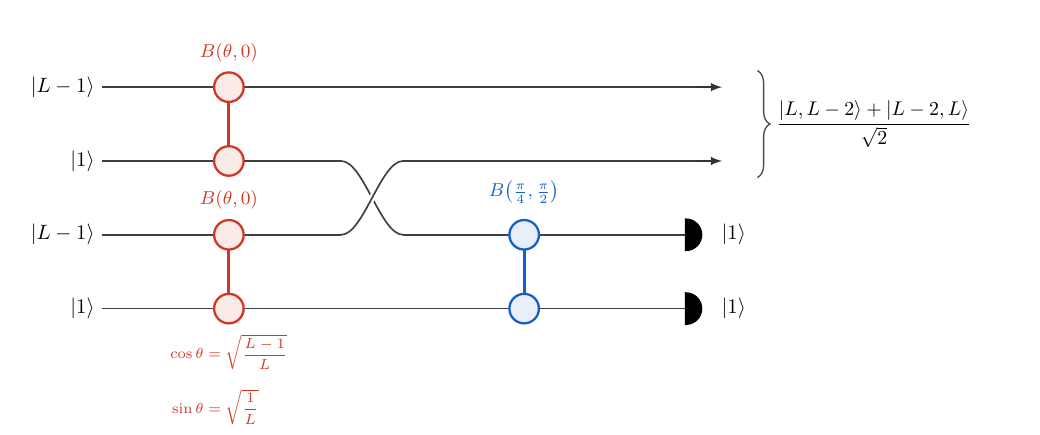}%
    }
    \caption{Explicit heralded construction for the $m,m'$-state family presented in Table~\ref{tab:additional_schemes}.}
    \label{fig:m_m'}
\end{figure}

\noindent
Finally, we consider the family presented in the last entry of
Table~\ref{tab:additional_schemes} and shown in
Fig.~\ref{fig:m_m'}.

\noindent
We denote the two target modes by $a_0,a_1$ and the two heralding
modes by $h_0,h_1$. The input state is
\begin{equation*}
    \ket{\psi_{\rm in}}
    =
    \ket{L-1,1,L-1,1}_{a_0,h_0,a_1,h_1},
\end{equation*}
which we regard as two identical blocks,
\begin{equation*}
    \ket{\psi_{\rm in}}
    =
    \ket{L-1,1}_{a_0,h_0}
    \otimes
    \ket{L-1,1}_{a_1,h_1}.
\end{equation*}
In each block $j\in\{0,1\}$, the mode $a_j$ is retained as a target
mode, while the mode $h_j$ is sent to the final heralding beam
splitter. The modes $a_j$ and $h_j$ are first mixed on an identical
beam splitter with
\begin{equation}
    t=\sqrt{\frac{L-1}{L}},
    \qquad
    r=\sqrt{\frac{1}{L}}.
    \label{eq:appendix_mmprime_bs}
\end{equation}

\noindent
We first analyze a generic block $(a_j,h_j)$. Its input state is
\begin{equation*}
    \ket{L-1,1}_{a_j,h_j}
    =
    \frac{
        \left(\hat a_j^\dagger\right)^{L-1}
        \hat h_j^\dagger
    }{\sqrt{(L-1)!}}
    \ket{\mathrm{vac}}_{a_j,h_j}.
\end{equation*}
The components relevant to the final
$\ket{1,1}_{h_0,h_1}$ heralding event are those containing zero, one,
or two photons in the heralding mode $h_j$ after the local beam
splitter.

The component with zero photons in $h_j$ is
\begin{equation*}
    \ket{L,0}_{a_j,h_j},
\end{equation*}
with amplitude
\begin{equation*}
    A_0
    =
    -\sqrt{L}\,r\,t^{L-1}
    =
    -t^{L-1}.
\end{equation*}
The component with two photons in $h_j$ is
\begin{equation*}
    \ket{L-2,2}_{a_j,h_j}.
\end{equation*}
Its amplitude is obtained by coherently combining the two
indistinguishable contributions that leave two photons in $h_j$:
\begin{align*}
    A_2
    &=
    \sqrt{\frac{2}{L-1}}
    \left[
        (L-1)t^{L-1}r
        -
        \binom{L-1}{2}t^{L-3}r^3
    \right]
    \nonumber\\
    &=
    \sqrt{\frac{2}{L-1}}\,
    (L-1)t^{L-3}r
    \left(
        t^2-\frac{L-2}{2}r^2
    \right).
\end{align*}
Using Eq.~\eqref{eq:appendix_mmprime_bs},
\begin{equation*}
    t^2-\frac{L-2}{2}r^2
    =
    \frac{1}{2},
\end{equation*}
and therefore
\begin{equation*}
    A_2
    =
    \frac{t^{L-2}}{\sqrt{2}}.
\end{equation*}

\noindent
After applying the two identical local beam splitters, the terms that
can produce the desired target components together with two total
heralding photons are
\begin{equation*}
    \ket{L,L-2}_{a_0,a_1}
    \ket{0,2}_{h_0,h_1}
\end{equation*}
and
\begin{equation*}
    \ket{L-2,L}_{a_0,a_1}
    \ket{2,0}_{h_0,h_1}.
\end{equation*}
Both terms have amplitude
\begin{equation*}
    A_0A_2
    =
    -\frac{t^{2L-3}}{\sqrt{2}}.
\end{equation*}

There is also a contribution containing one photon in each heralding
mode,
\begin{equation*}
    \ket{L-1,L-1}_{a_0,a_1}
    \ket{1,1}_{h_0,h_1},
\end{equation*}
which would leave the unwanted target state
$\ket{L-1,L-1}_{a_0,a_1}$. We remove this contribution by mixing the
heralding modes $h_0,h_1$ on a final $50{:}50$ beam splitter with
\begin{equation}
    t_h=\frac{1}{\sqrt{2}},
    \qquad
    r_h=\frac{i}{\sqrt{2}}.
    \label{eq:appendix_mmprime_final_bs}
\end{equation}
With this choice, an input
$\ket{1,1}_{h_0,h_1}$ has zero amplitude to be detected as
$\ket{1,1}_{h_0,h_1}$ at the output, owing to the usual two-photon
destructive interference. On the other hand,
\begin{equation*}
    \ket{2,0}_{h_0,h_1}
    \longrightarrow
    \frac{i}{\sqrt{2}}
    \ket{1,1}_{h_0,h_1}
    +\cdots,
\end{equation*}
and
\begin{equation*}
    \ket{0,2}_{h_0,h_1}
    \longrightarrow
    \frac{i}{\sqrt{2}}
    \ket{1,1}_{h_0,h_1}
    +\cdots.
\end{equation*}
Thus, both desired target components acquire the same heralding
amplitude.

Conditioned on detecting $\ket{1,1}_{h_0,h_1}$ at the two heralding
outputs, the unnormalized state of the target modes $a_0,a_1$ is
therefore
\begin{equation*}
    \ket{\psi_{\rm tar}}_{a_0,a_1}
    =
    -\frac{i\,t^{2L-3}}{2}
    \left(
        \ket{L,L-2}_{a_0,a_1}
        +
        \ket{L-2,L}_{a_0,a_1}
    \right).
\end{equation*}
After normalization, the heralded target state is
\begin{equation*}
    \frac{
        \ket{L,L-2}_{a_0,a_1}
        +
        \ket{L-2,L}_{a_0,a_1}
    }{\sqrt{2}},
\end{equation*}
up to an irrelevant global phase.

The success probability is the squared norm of the unnormalized
target state:
\begin{equation*}
    \psucc(L)
    =
    2
    \left|
        \frac{t^{2L-3}}{2}
    \right|^2
    =
    \frac{1}{2}t^{4L-6}.
\end{equation*}
Finally, since $t^2=(L-1)/L$, we obtain
\begin{equation*}
    \psucc(L)
    =
    \frac{1}{2}
    \left(
        1-\frac{1}{L}
    \right)^{2L-3}.
\end{equation*}

\end{document}